\documentclass[fleqn,usenatbib,useAMS]{mnras}
\usepackage{graphicx}
\usepackage{amsfonts}
\usepackage{amssymb}
\usepackage{amsmath}
\usepackage{color}

\newcommand{\HII}{H\,\small{II}\normalsize\,\,}

\title[A new perspective on Galactic magnetic fields]{A new perspective on turbulent Galactic magnetic fields through comparison of linear polarisation decomposition techniques}

\author[J.-F. Robitaille et al.]{J.-F. Robitaille$^{1}$\thanks{E-mail:
jean-francois.robitaille@manchester.ac.uk}, A. M. M. Scaife$^{1}$, E. Carretti$^{2,3}$, B. M. Gaensler$^{4}$, J. D. McEwen$^{5}$,
\newauthor B. Leistedt$^{6,7}$, M. Haverkorn$^{8}$, G. Bernardi$^{9,10}$, M. J. Kesteven$^{3}$, S. Poppi$^{2}$
\newauthor and L. Staveley-Smith$^{11,12}$\\
$^{1}$Jodrell Bank Centre for Astrophysics, School of Physics and Astronomy,
\\ The University of Manchester, Oxford Road, Manchester M13 9PL, UK\\
$^{2}$INAF Osservatorio Astronomico di Cagliari, Via della Scienza 5, 09047 Selargius (CA), Italy\\
$^{3}$CSIRO Astronomy and Space Science, PO Box 76, Epping, NSW 1710, Australia\\
$^{4}$Dunlap Institute for Astronomy and Astrophysics, The University of Toronto, Toronto, ON M5S 3H4, Canada\\
$^{5}$Mullard Space Science Laboratory (MSSL), University College London (UCL), Surrey RH5 6NT, UK\\
$^{6}$Department of Physics \& Astronomy, University College London, Gower Street, London WC1E 6BT, UK\\
$^{7}$Center for Cosmology and Particle Physics, Department of Physics, New York University, 726 Broadway, New York, NY 10003, USA\\
$^{8}$Department of Astrophysics/IMAPP, Radboud University, P.O. Box 9010, NL-6500 GL Nijmegen, The Netherlands\\
$^{9}$Harvard-Smithsonian Center for Astrophysics, 60 Garden Street, Cambridge, MA 02138, USA\\
$^{10}$SKA South Africa, 3rd Floor, The Park, Park Road, Pinelands 7405, South Africa\\
$^{11}$International Centre for Radio Astronomy Research, M468, University of Western Australia, Crawley WA 6009, Australia\\
$^{12}$ARC Centre of Excellence for All-sky Astrophysics (CAASTRO), Australia}

\begin{document}

\date{Accepted 2017 March 10}

\pagerange{\pageref{firstpage}--\pageref{lastpage}} \pubyear{2017}

\maketitle

\label{firstpage}

\begin{abstract}

\noindent We compare two rotationally invariant decomposition techniques on linear polarisation data: the spin-2 spherical harmonic decomposition in two opposite parities, the $E$- and $B$-mode, and the multiscale analysis of the gradient of linear polarisation, $|\nabla \bmath{P}|$. We demonstrate that both decompositions have similar properties in the image domain and the spatial frequency domain. They can be used as complementary tools for turbulence analysis of interstellar magnetic fields in order to develop a better understanding of the origin of energy sources for the turbulence, the origin of peculiar magnetic field structures and their underlying physics. We also introduce a new quantity $|\nabla EB|$ based on the $E$- and $B$-modes and we show that in the intermediate and small scales limit $|\nabla EB| \simeq |\nabla \bmath{P}|$. Analysis of the 2.3 GHz S-band Polarization All Sky Survey (S -PASS) shows many extended coherent filament-like features appearing as `double-jumps' in the $|\nabla \bmath{P}|$ map that are correlated with negative and positive filaments of $B$-type polarisation. These local asymmetries between the two polarisation types, $E$ and $B$, of the non-thermal Galactic synchrotron emission have an influence on the $E$- and $B$-mode power spectra analyses. The wavelet-based formalism of the polarisation gradient analysis allows us to locate the position of $E$- or $B$-mode features responsible for the local asymmetries between the two polarisation types. In analysed subregions, the perturbations of the magnetic field are trigged by star clusters associated with HII regions, the Orion-Eridanus superbubble and the North Polar Spur at low Galactic latitude.

\end{abstract}


\section{Introduction}\label{sec:intro}

Magnetic fields are omnipresent in the Milky Way and play a crucial role in many physical processes in the interstellar medium (ISM). Because interstellar matter is never completely neutral, magnetic fields are locked into a diffuse plasma and have a significant influence on the distribution of matter through the ISM. The interaction between the magnetic field lines and the gas induces an additional pressure in the ISM and can control the star formation process in the Galaxy \citep{2012ARA&A..50...29C}. Magnetic fields are also essential for the acceleration of charged particles leading to the propagation of cosmic rays. Unfortunately, magnetic fields cannot be observed directly. Since we know that they are acting on a broad range of spatial scales in the Galaxy, we need a tracer which allows us to connect the large-scale field, which is believed to be regular and following the morphology of Galactic arms, with the small-scale field, which is believed to be affected by turbulence and quasi-random. Diffuse synchrotron emission is present everywhere in the Galaxy and therefore is a powerful tracer of the magnetised turbulent non-thermal medium over a broad range of spatial scales. Many surveys of the Galactic polarised emission over a large portion of the sky and including large-scale structures have now been performed (e.g. The Canadian  Galactic Plane Survey (CGPS) \citep{2010A&A...520A..80L}, The Southern Galactic Plane Survey (SGPS) \citep{2006ApJS..167..230H}, The S-band Polarization All Sky Survey (S-PASS) (Carretti et al., in prep.; Carretti et al. \citeyear{2013Natur.493...66C}) or are on the way to being completed (e.g. The G-ALFA Continuum Transit Survey (GALFACTS) \citep{2010ASPC..438..402T}, The GaLactic and Extragalactic All-sky MWA survey (GLEAM) \citep{2015PASA...32...25W}). These surveys are giving us a new view of the Galactic magnetic field structure and are revealing its complexity at all spatial scales.

However to interpret the structures in such polarised emission data, we need robust analysing techniques. Stokes parameters $Q$ and $U$ are not invariant under arbitrary translations and rotations and choice of the local coordinate frame. Such shifts are common and can result, for example, from a smooth distribution of intervening polarised emission, a smooth uniform screen of foreground Faraday rotation, the effects of missing large-scale structure in an interferometric data set, or a combination of any of the three. Also, polarisation data are usually interpreted in terms of the amplitude, $|\bmath{P}|=\sqrt{Q^2+U^2}$, or the angle, $\theta=(1/2)\arctan(U/Q)$, of the polarisation vector\footnote{Strictly speaking, the linear polarisation is a tensor. The linearly polarised light is fully described by the rank-2 intensity tensor $I_{ij}$, where $Q=(I_{11}-I_{22})/4$ and $U=(I_{12}+I_{21})/4$ \citep{2004ASSL..307.....L}. However, the name vector is kept in this paper for simplicity. To avoid the $\pi$ ambiguity and to recover the full range of polarisation angles ($[-\pi/2,\pi/2]$), the two-argument function is used to define the polarisation angle as $\theta=(1/2)\arg(\bmath{P})=(1/2)$arctan($U,Q$). The orientation of the polarisation vector respects the IAU convention, which is $\theta=0^{\circ}$ toward Galactic north and $\theta$ increasing eastward.\label{footnote:atan2}} $\bmath{P}=Q+iU$ separately, which leads to an incomplete analysis of the magnetic field power spectrum in the ISM. 

In order to create two scalar rotationally invariant quantities, \citet{1997PhRvD..55.1830Z} introduced the spin-2 decomposition of the Stokes $Q$ and $U$ signal in a scalar and a pseudo-scalar quantity commonly called the $E$- and $B$-mode polarisation. This decomposition is mainly used for cosmic microwave background (CMB) analysis since density fluctuations of the primordial radiation produce $E$-type polarisation only. On the other hand, $B$-type polarisation can be produced by parity-odd components such as the presence of a stochastic background of gravitational waves \citep{1993PhRvL..71..324C,1997PhRvL..78.2058K,1997PhRvL..78.2054S,2010ASPC..438..276C}. By construction, the spin-2 decomposition of Stokes $Q$ and $U$ is non-local and thus sometimes makes its interpretation non-intuitive. In order to gain an intuition on the nature of the decomposition, \citet{2001PhRvD..64j3001Z} showed examples of how polarisation patterns inside filaments would create $E$- or $B$-type structures and how the emission of a supernova remnant (SNR) would look in these modes. These kind of polarisation patterns have been already detected in the magnetised filamentary structure of infrared dust emission \citep{2016A&A...586A.141P} and they are believed to be responsible for the $E$--$B$ asymmetry reported in the power spectra analysis of the \textit{Planck} 353 GHz polarisation maps \citep{2016A&A...586A.133P, 2016A&A...586A.141P}. \citet{2015arXiv150201588P} also reported an $E$--$B$ asymmetry for the synchrotron emission at 30 GHz.

\citet{2011Natur.478..214G} also introduced a new quantity invariant under arbitrary translations and rotations of the $Q$--$U$ plane, the gradient of linear polarisation. It is defined as,

\begin{equation}
|\nabla\bmath{P}| = \sqrt{\left(\frac{\partial Q}{\partial x}\right)^2+\left(\frac{\partial U}{\partial x}\right)^2 + \left(\frac{\partial Q}{\partial y}\right)^2+\left(\frac{\partial U}{\partial y}\right)^2}.
\label{eq:gradP}
\end{equation}

\noindent Developed for the analysis of radio synchrotron emission, the gradient acts as an edge detector in a polarisation map and highlights areas of sharp change in the magnetic field and/or the free-electron density. \citet{2015MNRAS.451..372R} demonstrated that the gradient of $\bmath{P}$ analysis can be extended to multiple spatial scales, allowing one to calculate the power spectrum of the gradient of $\bmath{P}$ and to give a complete measure of the magnetic field fluctuations.

The aim of this paper is to underline the interesting relationships that exist between both methods and demonstrate that analyses of local variations in both decompositions can be used as complementary tools in order to develop a better understanding of the origin of energy sources for the turbulent magnetic field, the origin of peculiar magnetic field structures and their underlying physics. The paper is organised as follows: the data set is presented in section \ref{sec:observation}, a brief review of the multiscale analysis of the gradient of linear polarisation is given in section \ref{sec:multiscale}, a comparison between $|\nabla \bmath{P}|$ and the spin-2 decomposition is done in section \ref{sec:gardPvsSpin2}, results and analysis of particular regions are presented in \ref{sec:results}, power spectra analysis and the influence of $E$- and $B$-modes asymmetries are analysed in section \ref{sec:pow_spec} and a discussion and a conclusion on the main results of this paper are in sections \ref{sec:discussion} and \ref{sec:conclusion}.

\section{Observations}\label{sec:observation}

Our analysis is performed on the S-band Polarization All Sky Survey (S-PASS). S-PASS is a single-dish polarimetric survey of the entire southern sky at 2.3 GHz, performed with the Parkes 64 m Radio Telescope and its S-band receiver with beam width FWHM=8.9 arcmin (Carretti et al., in prep. and \citet{2013Natur.493...66C}, see also \citet{2010MNRAS.405.1670C} and \citet{2014A&A...566A...5I}). Final maps were convolved with a Gaussian window of FWHM = 6 arcmin to a final resolution of 10.75 arcmin. We use the Stokes $Q$ and $U$ maps produced by this survey to calculate the multiscale gradient of $\bmath{P}$ and the spin-2 decomposition.

\section{Multiscale analysis of the gradient of linear polarisation} \label{sec:multiscale}

The development and first application of the multiscale $|\nabla \bmath{P}|$ method was presented by \citet{2015MNRAS.451..372R}. Here we review the basic principles of the method and expand the original description to provide the reader with a more complete understanding of the work presented in this paper.

\citet{2015MNRAS.451..372R} showed that the spatial gradient of linear polarisation can be calculated using a wavelet formalism. This formalism has properties similar to the $\Delta$-variance analysis used to characterise multiscale structures induced by turbulence in molecular clouds \citep{1998A&A...336..697S, 2001A&A...366..636B, 2008A&A...485..917O, 2008A&A...485..719O, 2011A&A...529A...1S,2014A&A...568A..98A}. The technique also shares similarities with the wavelet transform modulus maxima (WTMM) method used to characterise the multifractal nature of a surface or to find singularities in a signal \citep{2000EPJB...15..567A, 2006ApJS..165..512K, 2010ApJ...717..995K}. By taking advantage of the smooth filter shape of wavelets localised in the spatial and the frequency domain, this wavelet formalism allows us to expand the analysis of the polarisation gradient to multiple spatial scales and, importantly, to derive the power spectrum of that quantity.

In order to perform a multiscale $|\nabla \bmath{P}|$ analysis, the continuous wavelet transform of Stokes parameters $Q$ and $U$ must be calculated at different spatial scales. The continuous wavelet transform is defined as 

\begin{equation}
\tilde{f}(l,\bmath{x}) =
\begin{cases}
\tilde{f}_1 = l^{-1} \int \psi_1[l^{-1}(\bmath{x}-\bmath{x'})]f(\bmath{x'}) d^2\bmath{x'}\\
\tilde{f}_2 = l^{-1} \int \psi_2[l^{-1}(\bmath{x}-\bmath{x'})]f(\bmath{x'}) d^2\bmath{x'},
\end{cases}
\label{eq:DoG_transforms}
\end{equation}

\noindent where $\psi_1(\bmath{x})$ and $\psi_2(\bmath{x})$ are the wavelet functions and $l$ is the wavelet scaling factor. The function $f(\bmath{x})$ represents the map of Stokes $Q$ or $U$. The wavelet functions are defined as

\begin{equation}
\begin{split}
\psi_1(x,y) &= \partial^m \phi(x,y)/ \partial x^m,\\
\psi_2(x,y) &= \partial^m \phi(x,y)/ \partial y^m,
\end{split}
\label{eq:Gauss_derivative}
\end{equation}

\noindent where the function $\phi$ is a Gaussian distribution,

\begin{equation}
\phi(x,y) = \frac{1}{2\pi } e^{\frac{-(x^2+y^2)}{2}}.
\label{eq:Gaussian}
\end{equation}

\begin{figure}
\centering
\includegraphics[]{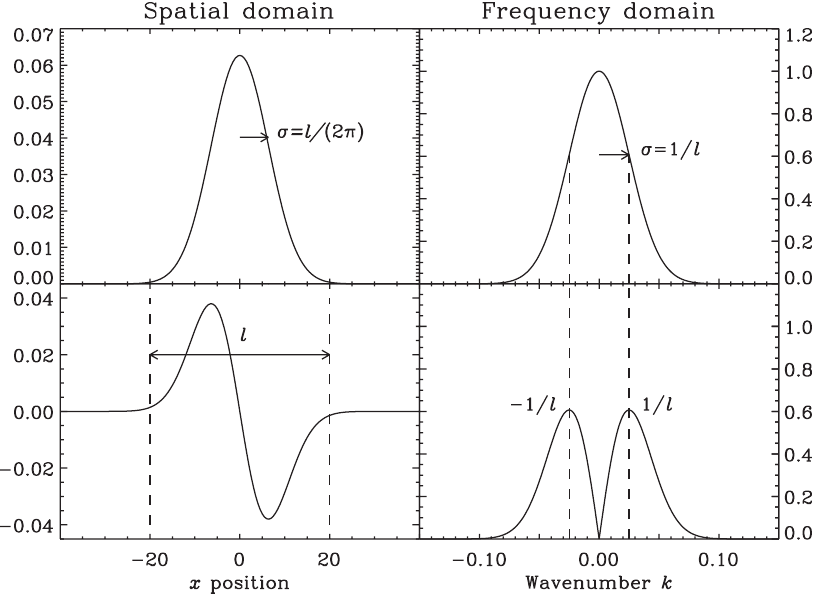}
\caption{Top left: a one-dimensional Gaussian distribution with $\sigma=l_{\textrm{F}}$. Top right: the amplitude of the Fourier transform of the Gaussian distribution. Bottom left: the first derivative of $\phi(x)$, i.e. the DoG wavelet. Bottom right: the amplitude of the Fourier transform of the DoG wavelet.}
\label{fig:gauss_dog}
\end{figure}

\noindent Even values of $m$ produce symmetric wavelet functions and and odd values produce asymmetric wavelet functions. For $m=1$, equation (\ref{eq:DoG_transforms}) is equivalent to the calculation of the gradient of $f(\bmath{x})$ smoothed by the Gaussian kernel $\phi$:

\begin{equation}
\tilde{f}(l,\bmath{x}) = l^{-1} \nabla \{\phi(l^{-1}\bmath{x}) \otimes f(\bmath{x}) \},
\label{eq:grad_Gauss}
\end{equation}

\noindent where $\otimes$ is the convolution operation\footnote{The relation in equation (\ref{eq:grad_Gauss}) differs to that of equation (5) in \citet{2015MNRAS.451..372R}. This modification is consistent with the normalisation factor $l^{-1}$ in equation (\ref{eq:DoG_transforms}) in the present paper. It is worth noting that this definition differs from the one initially proposed by \citet{2000EPJB...15..567A}.}. Figure \ref{fig:gauss_dog} shows how the derivative of the Gaussian function changes the properties of the filtering function in the frequency domain. The Fourier transform of a Gaussian function conserves the same shape in the frequency domain and acts as a low-pass frequency filter, damping the high frequency content of the signal. The derivative of a function in the frequency domain is defined as $\hat{\phi}^m(k) = (2\pi ik)^m\hat{\phi}(k)$, where $\hat{\phi}$ represents the Fourier transform of $\phi$, $m$ the order of the derivative and $k$ the wavenumber. Following this definition, the derivative in the frequency domain becomes a weighting function applied to the amplitude of the Gaussian distribution, which, according to Fig. \ref{fig:gauss_dog}, modifies the function and produces a band-pass frequency filter. In the spatial domain, the Derivative of Gaussian, or DoG wavelet, becomes an oscillatory function which satisfies all the properties of a wavelet function.

As described by \citet{2015MNRAS.451..372R}, the wavelet scaling factor $l$ of equations (\ref{eq:DoG_transforms}) and (\ref{eq:grad_Gauss}) is substituted by $l_{\textrm{F}} = \sqrt{m}\cdot l\cdot(2\pi)^{-1}$ in order to facilitate comparison of the scaling factor $l$ with the Fourier wavenumber $k=1/l$. Figure \ref{fig:gauss_dog} shows the size and the position of a one-dimensional Gaussian distribution

\begin{equation}
l_{\textrm{F}}^{-1}\phi(l_{\textrm{F}}^{-1}x) = \frac{1}{l_{\textrm{F}}\sqrt{2\pi}} e^{-0.5(x/l_{\textrm{F}})^2},
\label{eq:Gauss_1D}
\end{equation}

\noindent with a standard deviation of $l_{\textrm{F}}$ compared with the wavelet function

\begin{equation}
l_{\textrm{F}}^{-1}\psi(l_{\textrm{F}}^{-1}x) = \frac{-x}{l_{\textrm{F}}^2\sqrt{2\pi}}e^{-0.5(x/l_{\textrm{F}})^2},
\label{eq:DoG_1D}
\end{equation}

\noindent which corresponds to the first derivative of a Gaussian multiplied by the normalisation factor $l_{\textrm{F}}^{-1}$ as in equation (\ref{eq:DoG_transforms}). Figure \ref{fig:gauss_dog} shows that this substitution allows one to relate the angular size $l$ of the wavelet function in the spatial domain with the position $1/l$ of the band-pass filter in the frequency domain.

The new definition of the multiscale polarisation gradient is

\begin{equation}
|\nabla \tilde{\bmath{P}}(l,\bmath{x})| = \sqrt{ |\nabla \tilde{Q}(l,\bmath{x})|^2 + |\nabla \tilde{U}(l,\bmath{x})|^2},
\label{eq:scaled_gradient_P}
\end{equation}

\noindent where, referring to equation (\ref{eq:DoG_transforms}),

\begin{equation}
\begin{split}
|\nabla \tilde{Q}(l,\bmath{x})| & = \sqrt{ |\tilde{Q}_{1}(l,\bmath{x})|^2 + |\tilde{Q}_{2}(l,\bmath{x})|^2},\\
\\
|\nabla \tilde{U}(l,\bmath{x})| & = \sqrt{ |\tilde{U}_{1}(l,\bmath{x})|^2 + |\tilde{U}_{2}(l,\bmath{x})|^2}.\\
\end{split}
\label{eq:Stokes_amplitude}
\end{equation}

In this paper, the multiscale analysis of $|\nabla \bmath{P}|$ is considered for the complete southern hemisphere as well as for subregions of the S-PASS survey. The analysis of the normalised spatial gradient of linear polarisation, $|\nabla \bmath{P}|/|\bmath{P}|$, has already been applied by \citet{2014A&A...566A...5I} on the S-PASS survey at its original resolution. They analysed in detail the morphology of several objects, such as \HII regions and SNRs, and compared also the high-order moments of observed $|\nabla \bmath{P}|$ structures with $|\nabla \bmath{P}|$ structures of real MHD simulations realised at different sonic Mach numbers. Here, we extend the analysis on many spatial scales and compare the filamentary structures of the $|\nabla \bmath{P}|$ images with extended linear polarisation patterns. The turbulent component is analysed through the $|\nabla \bmath{P}|$ power spectrum.

Analysis of the southern sky hemisphere is done by projecting the map on \texttt{HEALPix}\footnote{\url{http://healpix.jpl.nasa.gov}} pixelisation format with a $N_{\rm{side}}$ resolution of 2048. Each pixel produced with the \texttt{HEALPix} tessellations of the sphere covers the same surface area and thus avoids the `compression effect' on structures located at the edge of the maps. On the sphere, the multiscale polarisation gradient is calculated using equation (\ref{eq:grad_Gauss}). The \texttt{HEALPix} map is first smoothed using the \texttt{ismoothing HEALPix} procedure and then the derivative is calculated using outputs of the \texttt{isynfast} \texttt{IDL} procedure. The derivation of the DoG wavelet on the sphere and its convolution to a spherical map would also be possible, however such calculation is beyond the scope of this paper.

\section{Comparison between $|\nabla \bmath{P}|$ and the spin-2 decomposition} \label{sec:gardPvsSpin2}

The $|\nabla \bmath{P}|$ method was originally proposed by \citet{2011Natur.478..214G} as a quantity unaffected by arbitrary translations, or missing zero-offset, or rotations of the $Q$--$U$ plane unlike quantities such as the polarisation amplitude and polarisation angle which are not preserved under arbitrary translations and rotations. In the image domain, the invariance property of $|\nabla \bmath{P}|$, associated with the shift of the polarisation vector caused, for example, by a foreground Faraday screen, can be interpreted differently. As demonstrated in the previous section, the derivative of a map at a certain resolution produces a band-pass filter in spatial scale. Consequently, this filtering operation prevents any translation or rotation of the polarisation vector in the image plane that would be caused by the superposition of larger fluctuations on the map.

Another way to define a rotationally invariant quantity from the Stokes parameters $Q$ and $U$ is with the spin $\pm2$ spherical harmonic\footnote{A function $\zeta(\theta,\phi)$, defined on the sphere, has a spin weight $s$ if it transforms as $\zeta'=e^{si\psi}\zeta$ under rotation, where $\psi$ is the angle of rotation \citep{1966JMP.....7..863N}.} decomposition in two opposite parities, the magnetic-type parity ($B$-modes) and the electric-type parity ($E$-modes) \citep{1997PhRvD..55.1830Z}. The $E$-mode, like the polarised intensity $|\bmath{P}|$, is a scalar quantity and the $B$-mode is a pseudo-scalar quantity. This means that their rotationally invariant power spectra can be easily calculated from their expansion in spherical harmonics:

\begin{equation}
C_{X \ell} = \frac{1}{2\ell+1}\sum_{m}|a_{X,\ell m}|^2,
\label{eq:Sphe_pow_spec}
\end{equation}

\noindent with $X = E$ or $B$ and

\begin{equation}
X(\bmath{\hat{n}}) = \sum_{\ell m} a_{X,\ell m}Y_{\ell m}(\bmath{\hat{n}}),
\label{eq:EB_sphe_harm}
\end{equation}

\noindent where $Y_{\ell m}(\theta, \phi)$ are the spherical harmonics and $a_{X,\ell m}$ are the spherical harmonic coefficients. The spherical harmonic expansion is the most natural way to describe the $E$- and  $B$-mode decomposition. However, in the small-scale limit, this decomposition can also be done on the plane using a standard Fourier transform, see \citet{1997ApJ...482....6S} and \citet{2002ApJ...579..607T} for more details.

The Stokes parameters $Q$ and $U$ and their transformation under rotation by an angle $\psi$ is defined as \citep{1966raas.book.....K}

\begin{equation}
\begin{array}{l}

Q' = Q\cos 2\psi + U\sin2\psi,\\
\\
U' = -Q\sin 2\psi + U\cos2\psi.

\end{array}
\label{eq:Stokes_rotation}
\end{equation}

\noindent This behaviour means that the Stokes parameters $Q$ and $U$ are coordinate dependent. For example, if $U=0$ in a particular coordinate system and one rotates this coordinate system, the rotation will change the value of $Q$ and add a $U$ component. This is why the calculation of a rotationally invariant power spectrum of $Q$ and $U$ is complicated, since, for each wavenumber, $Q$ and $U$ have to be rotated to a common frame before the superposition can be done \citep{1997PhRvD..55.1830Z}.

Following the definition in equations (\ref{eq:Stokes_rotation}), Stokes $Q$ and $U$ can also be defined as a spin $\pm2$ signal, $(Q\pm iU)'(\bmath{\hat{n}}) = \exp(\mp2i\psi)(Q\pm iU)(\bmath{\hat{n}})$. The transformation of the spin $\pm2$ signal into the scalar $E$-mode and pseudo-scalar $B$-mode is done essentially by calculating the linear combination of the spherical harmonic coefficients of both spin signals \citep{1997PhRvD..55.1830Z}:

\begin{equation}
\begin{array}{l}

a_{E,\ell m} = -(a_{+2,\ell m} + a_{-2,\ell m})/2,\\
\\
a_{B,\ell m} = i(a_{+2,\ell m} - a_{-2,\ell m})/2,

\end{array}
\label{eq:EB_linear_combination}
\end{equation}

\noindent where

\begin{equation}
a_{\pm2,\ell m} = \int d\Omega _{\pm2}Y^*_{\ell m}(\bmath{\hat{n}})(Q\pm iU)(\bmath{\hat{n}}),
\label{eq:spin_sphe_harm}
\end{equation}

\noindent and $_{\pm2}Y_{\ell m}(\bmath{\hat{n}})$ are spin-weighted spherical harmonics associated with spin $\pm2$ signals. The spin-2 spherical harmonics ($_{\pm2}Y_{\ell m}$) in equation (\ref{eq:spin_sphe_harm}) can also be written in terms of spin-0 spherical harmonics ($Y_{\ell m}$) using the spin raising and lowering operators, $\eth$ and $\overline{\eth}$ \citep{1966JMP.....7..863N}:

\begin{equation}
\begin{split}
_{2}Y_{\ell m} &= [(\ell-2)!/(\ell+2)!)]^{1/2}\eth\eth Y_{\ell m},\\
_{-2}Y_{\ell m} &= [(\ell-2)!/(\ell+2)!)]^{1/2}\overline{\eth}\overline{\eth} Y_{\ell m}.
\end{split}
\label{eq:spin_harmonics}
\end{equation}

\noindent In spherical coordinates, these operators appear as differential operators applied on both angular coordinates,

\begin{equation}
\begin{split}
\eth &\equiv -\sin^s\theta\left(\frac{\partial}{\partial\theta}+\frac{i}{\sin\theta}\frac{\partial}{\partial\varphi}\right)\sin^{-s}\theta,\\
\overline{\eth} &\equiv -\sin^{-s}\theta\left(\frac{\partial}{\partial\theta}-\frac{i}{\sin\theta}\frac{\partial}{\partial\varphi}\right)\sin^{s}\theta,
\end{split}
\label{eq:spin_operators}
\end{equation}

\noindent Using equations (\ref{eq:EB_linear_combination}) and (\ref{eq:spin_harmonics}), one can derive the relation in the real space between the spin-2 polarisation signal and the $E$- and $B$-mode as \citep{2003PhRvD..67b3501B}

\begin{equation}
\begin{split}
Q + iU &= \eth\eth (\Psi_E + i\Psi_B),\\
Q - iU &= \overline{\eth}\overline{\eth} (\Psi_E - i\Psi_B),
\end{split}
\label{eq:QU-EB_real_space}
\end{equation}

\noindent and inversely,

\begin{equation}
\begin{split}
\overline{\eth}\overline{\eth} \eth\eth\Psi_E &\equiv \frac{1}{2}[\overline{\eth}\overline{\eth}(Q+iU)+\eth\eth(Q-iU)],\\
\overline{\eth}\overline{\eth} \eth\eth\Psi_B &\equiv \frac{i}{2}[\overline{\eth}\overline{\eth}(Q+iU)-\eth\eth(Q-iU)],\\		      
\end{split}
\label{eq:EB_tilde}
\end{equation}

\noindent where,

\begin{equation}
\begin{split}
\Psi_E &= -\sum_{\ell m}\left[\frac{(\ell-2)!}{(\ell+2)!}\right]^{1/2} a_{E, \ell m} Y_{\ell m},\\
\Psi_B &= -\sum_{\ell m}\left[\frac{(\ell-2)!}{(\ell+2)!}\right]^{1/2} a_{B, \ell m} Y_{\ell m}.
\end{split}
\label{eq:EB_potentials}
\end{equation}

\noindent The quantities $\Psi_E$ and $\Psi_B$ are defined by \citet{2003PhRvD..67b3501B} as the scalar and pseudo-scalar `potentials' of $E$ and $B$ respectively. They are useful because they allow us to define the spin-2 signal $Q\pm iU$ in terms of the spin-0 quantities $\Psi_E$ and $\Psi_B$ using only the spin raising and lowering operators, $\eth$ and $\overline{\eth}$. In equations (\ref{eq:QU-EB_real_space}), they are respectively raising the spin from 0 to 2 and lowering the spin from 0 to $-2$.

In the flat-sky approximation, the left hand sides of equations (\ref{eq:EB_tilde}) are equivalent to $\nabla^2E$ and $\nabla^2B$ \citep{2003PhRvD..67b3501B}\footnote{\citet{1997PhRvD..55.1830Z} defined $\overline{\eth}\overline{\eth} \eth\eth\Psi_E$ and $\overline{\eth}\overline{\eth} \eth\eth\Psi_B$ as $\tilde{E}$ and $\tilde{B}$. In this paper, we keep the notation $\overline{\eth}^2 \eth^2\Psi_E$ and $\overline{\eth}^2 \eth^2\Psi_B$ to ovoid any confusion with our notation in equation (\ref{eq:DoG_transforms}) associated with the DoG wavelet transform.}. From this point of view, $\overline{\eth}^2 \eth^2\Psi_E$ and $\overline{\eth}^2 \eth^2\Psi_B$ share similarities with the gradient of the linear polarisation vector of equation (\ref{eq:gradP}). Assuming that scalar fields $E$ and $B$ are affected by a Gaussian window function, corresponding to the Parkes telescope beam in the case of S-PASS, the rotationally invariant quantities $\overline{\eth}^2 \eth^2\Psi_E$ and $\overline{\eth}^2 \eth^2\Psi_B$, in the planar approximation can be easily derived in a manner comparable to the wavelet approach described in equation (\ref{eq:grad_Gauss}):

\begin{equation}
\begin{split}
\overline{\eth}^2 \eth^2\Psi_E(l,\bmath{x}) &= l^{-1} \nabla^2 \{\phi(l^{-1}\bmath{x}) \otimes E(\bmath{x}) \},\\
\overline{\eth}^2 \eth^2\Psi_B(l,\bmath{x}) &= l^{-1} \nabla^2 \{\phi(l^{-1}\bmath{x}) \otimes B(\bmath{x}) \},
\end{split}
\label{eq:laplace_Gauss}
\end{equation}

\begin{figure}
\centering
\includegraphics[]{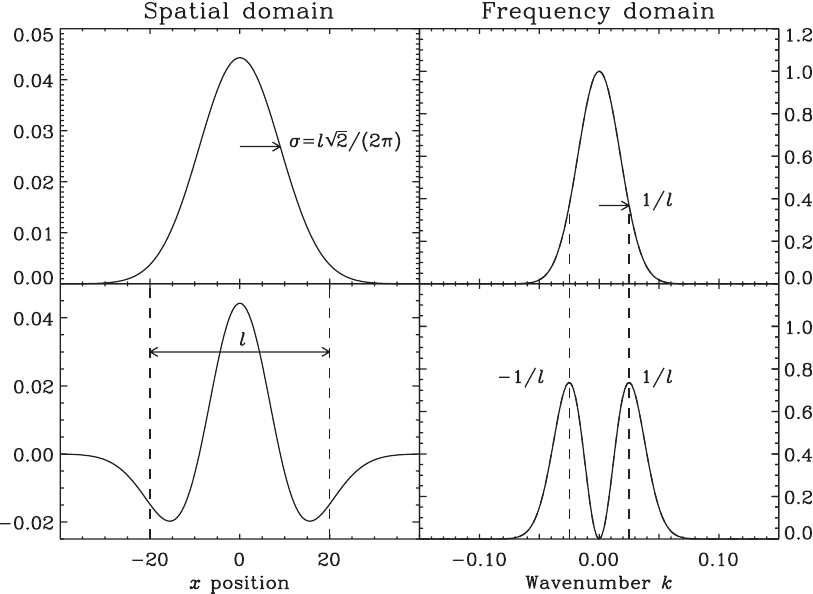}
\caption{Top left: the one-dimension Gaussian kernel $\phi(x)$ with $\sigma=l_{\textrm{F}}$. Top right: the Fourier transform of $\phi(x)$. Bottom left: the second derivative of $\phi(x)$. Bottom right: the Fourier transform of the DoG wavelet. Compared to the first derivative of $\phi(x)$ (see Fig. \ref{fig:gauss_dog}), it is interesting to note that the second derivative, using the same scaling factor $l$, has a better resolution in the Fourier domain.}
\label{fig:gauss_laplacian}
\end{figure}
\begin{figure}
\centering
\includegraphics[]{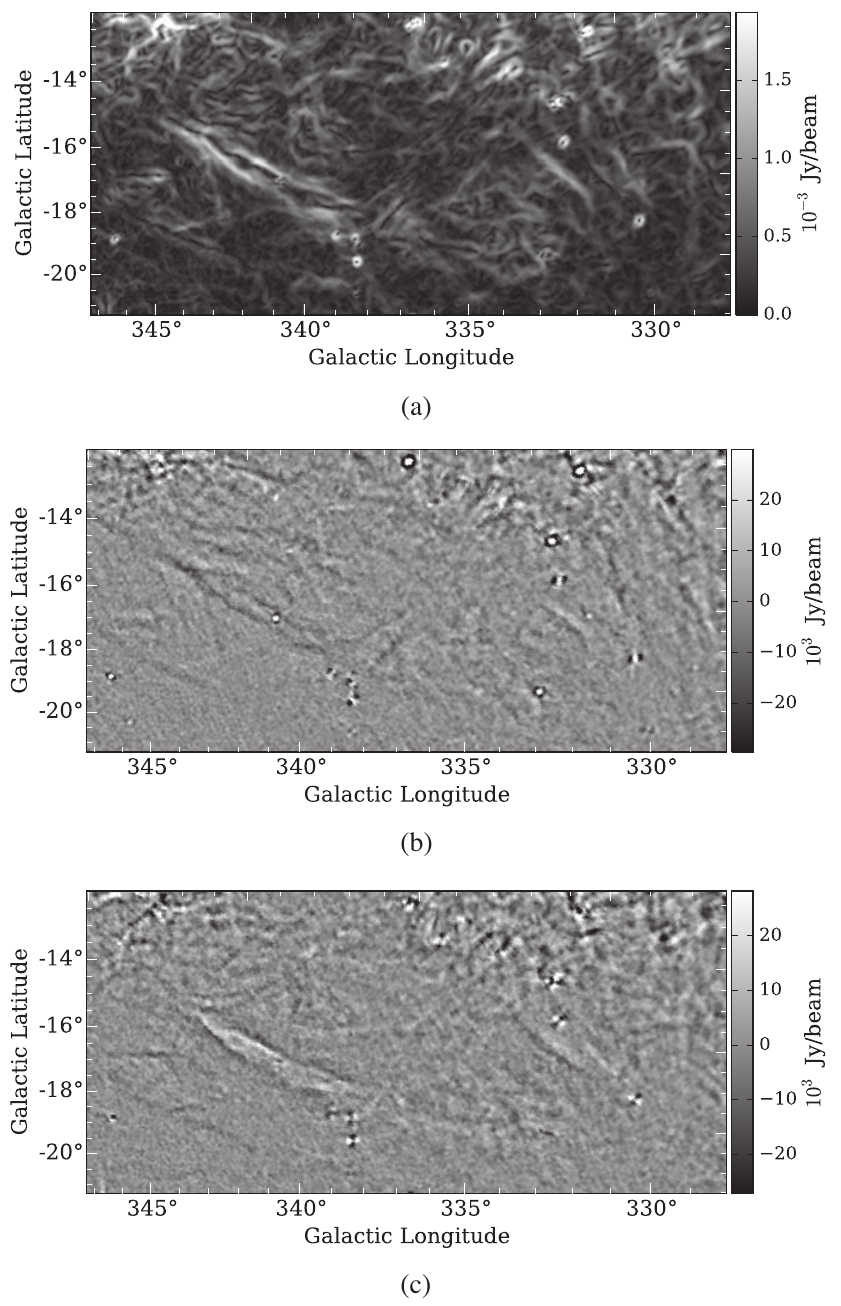}
\caption{(a) The $|\nabla\bmath{P}|$ map at an angular scale of $\sim$20 arcmin including the $|\nabla \textrm{RM}|$ feature B discussed by \citet{2014A&A...566A...5I}. (b) and (c) are respectively the $\overline{\eth}^2 \eth^2\Psi_E$ and $\overline{\eth}^2 \eth^2\Psi_B$ maps.}
\label{fig:gradPvstildeEB}
\end{figure}


\noindent where $E$ and $B$ are simply the spherical harmonic synthesis of $a_{E,\ell m}$ and $a_{B,\ell m}$ defined in equation (\ref{eq:EB_linear_combination}). Figure \ref{fig:gauss_laplacian} shows similar graphs to those in Fig. \ref{fig:gauss_dog}, but now with the second derivative applied to a Gaussian kernel. The same definition of the scaling factor $l$ is used. The second derivative of a Gaussian produces a symmetric function commonly known as the Mexican Hat. Its band-pass filter function in the Fourier domain is similar to that of the DoG wavelet, but since the second derivative of a Gaussian has more vanishing moments, the filter is more localised in the frequency domain. A comparison between $|\nabla\bmath{P}|$, $\overline{\eth}^2 \eth^2\Psi_E$ and $\overline{\eth}^2 \eth^2\Psi_B$ is shown in Fig. \ref{fig:gradPvstildeEB} using data from the S-PASS survey. The region includes the gradient of the rotation measure, $|\nabla \textrm{RM}|$, labelled `B' discussed by \citet{2014A&A...566A...5I}. In this paper, in order to experiment more easily with the spin operators, the HEALpix sampling is first converted in the \textit{McEwen-Wiaux} sampling presented by \citet{2011ITSP...59.5876M} on which we performed the spin spherical harmonic transform of $(Q \pm iU)$ using the \texttt{SSHT} code\footnote{\url{http://www.jasonmcewen.org}}, thereby dealing properly with the spin $\pm2$ complex signal. The code allowed us to compute directly equations (\ref{eq:EB_linear_combination}) and to perform the spin-0 spherical harmonic synthesis of the $E$- and $B$-mode spherical harmonic coefficients.

Like the $|\nabla\bmath{P}|$ analysis applied at the original map resolution, Fig. \ref{fig:gradPvstildeEB} shows that $\overline{\eth}^2 \eth^2\Psi_E$ and $\overline{\eth}^2 \eth^2\Psi_B$ maps display only the small-scale features, unless different scaling factors $l$ are used as demonstrated in section \ref{sec:multiscale}. However, the $|\nabla\bmath{P}|$ map traces more successfully sharp spatial changes in the polarisation vector than the $\overline{\eth}^2 \eth^2\Psi_E$ and $\overline{\eth}^2 \eth^2\Psi_B$ maps. This property can be attributed to the shape of the filters in the spatial domain and the fact that we are measuring the amplitude of the gradient.

The power spectra of $E$ and $B$ calculated from the linear combination in equation (\ref{eq:EB_linear_combination}) are by definition rotationally invariant since the two quantities are not coordinate dependent. Although, the $\overline{\eth}^2 \eth^2\Psi_E$ and $\overline{\eth}^2 \eth^2\Psi_B$ maps are rotationally invariant in the image domain for reasons similar to $|\nabla\bmath{P}|$. Like the polarisation gradient, the differential operators $\eth$ and $\overline{\eth}$ isolate the small-scale variations in the signal, which make the quantities $\overline{\eth}^2 \eth^2\Psi_E$ and $\overline{\eth}^2 \eth^2\Psi_B$ not sensitive to an intervening smooth Faraday screen. On the other hand, it is important to note that $E$ and $B$ maps are fundamentally different from the conventional $|\bmath{P}|$ or polarisation angle map. Without applying the differential operators $\eth$ and $\overline{\eth}$, $E$ and $B$ maps represent, respectively, a spin-0 scalar and pseudo-scalar field, which trace patterns in fluctuations occurring in $|\bmath{P}|$ and in the polarisation angle simultaneously. Furthermore, the scalar field $E$ remains unchanged under parity transformation while pseudo-scalar field $B$ changes sign. According to \citet{2001PhRvD..64j3001Z}, if the polarisation vectors are aligned or are perpendicular to the direction over which the magnitude of the polarisation is changing, the $E$-mode will dominate for that region. On the other hand, the $B$-mode dominates when the polarisation vectors are oriented at approximately $45^{\circ}$ to the change of magnitude.

\begin{figure}
\centering
\includegraphics[]{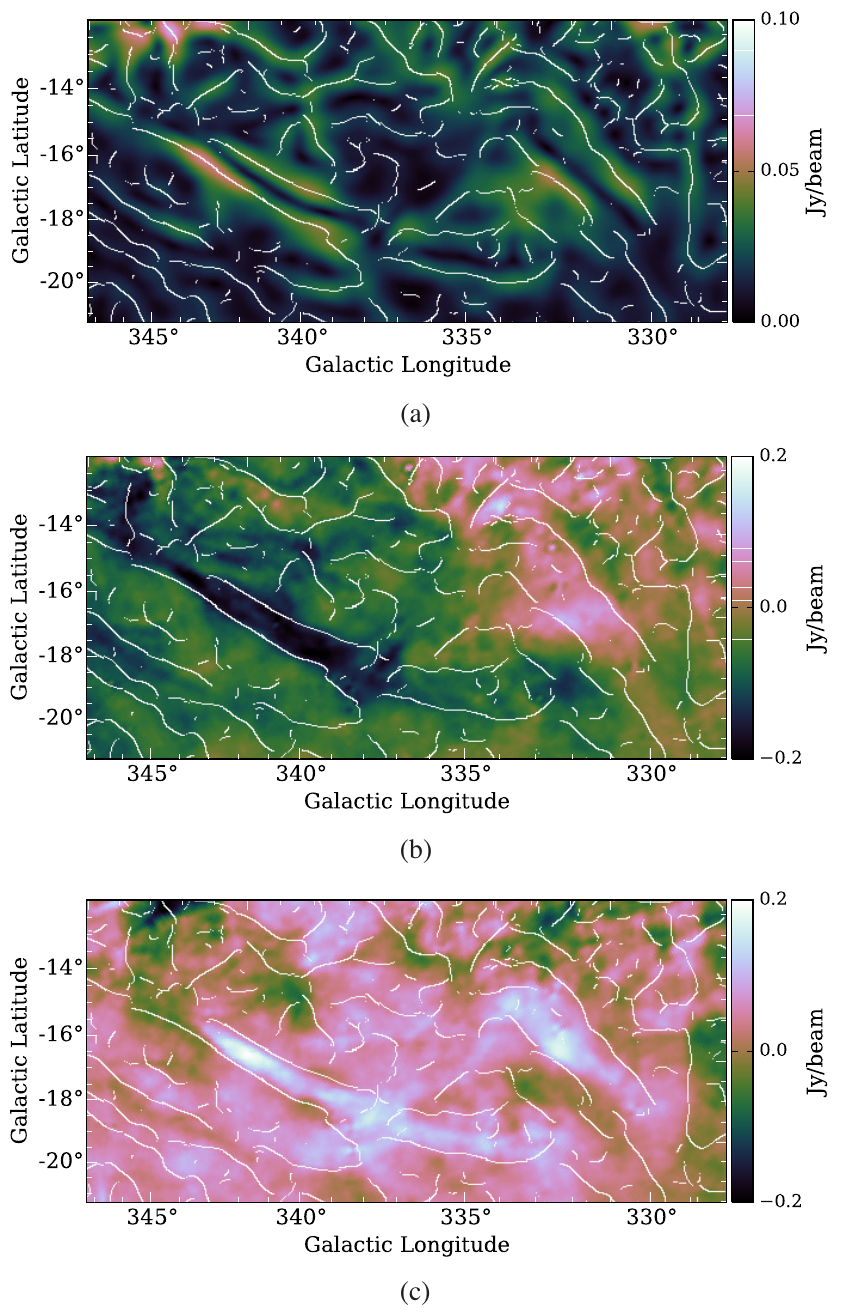}
\caption{(a) The polarisation gradient at an angular scale of $\sim 161$ arcmin. (b) and (c) represent respectively the $E$- and $B$-mode decomposition without the differential operators $\eth$ and $\overline{\eth}$. The white lines trace the $|\nabla\bmath{P}|$ maxima chains at the angular scale of $\sim 161$ arcmin.}
\label{fig:gradPvsEB}
\end{figure}

In this paper, we propose to examine different patterns in the $E$- and $B$-mode decomposition to study peculiar features seen in the polarised synchrotron emission. This analysis can be considered as complementary to the multiscale analysis of $|\nabla\bmath{P}|$ in order to develop a better understanding of the origin of these structures. As described by \citet{2014A&A...566A...5I}, gradients in the polarisation vector can be caused by fluctuations in different quantities and it is also true for $E$- and $B$-mode fluctuations in polarised synchrotron emission. For the analysis of the S-PASS data at 2.3 GHz, the synchrotron emission is likely affected by Faraday rotation along the line of sight. Thus, $|\nabla\bmath{P}|$ as well as $E$- and $B$-mode fluctuations represent a mix of RM fluctuations induced by the turbulent magneto-ionic medium and fluctuations in the polarised synchrotron foreground/background emission.

\begin{figure*}
\centering
\includegraphics[]{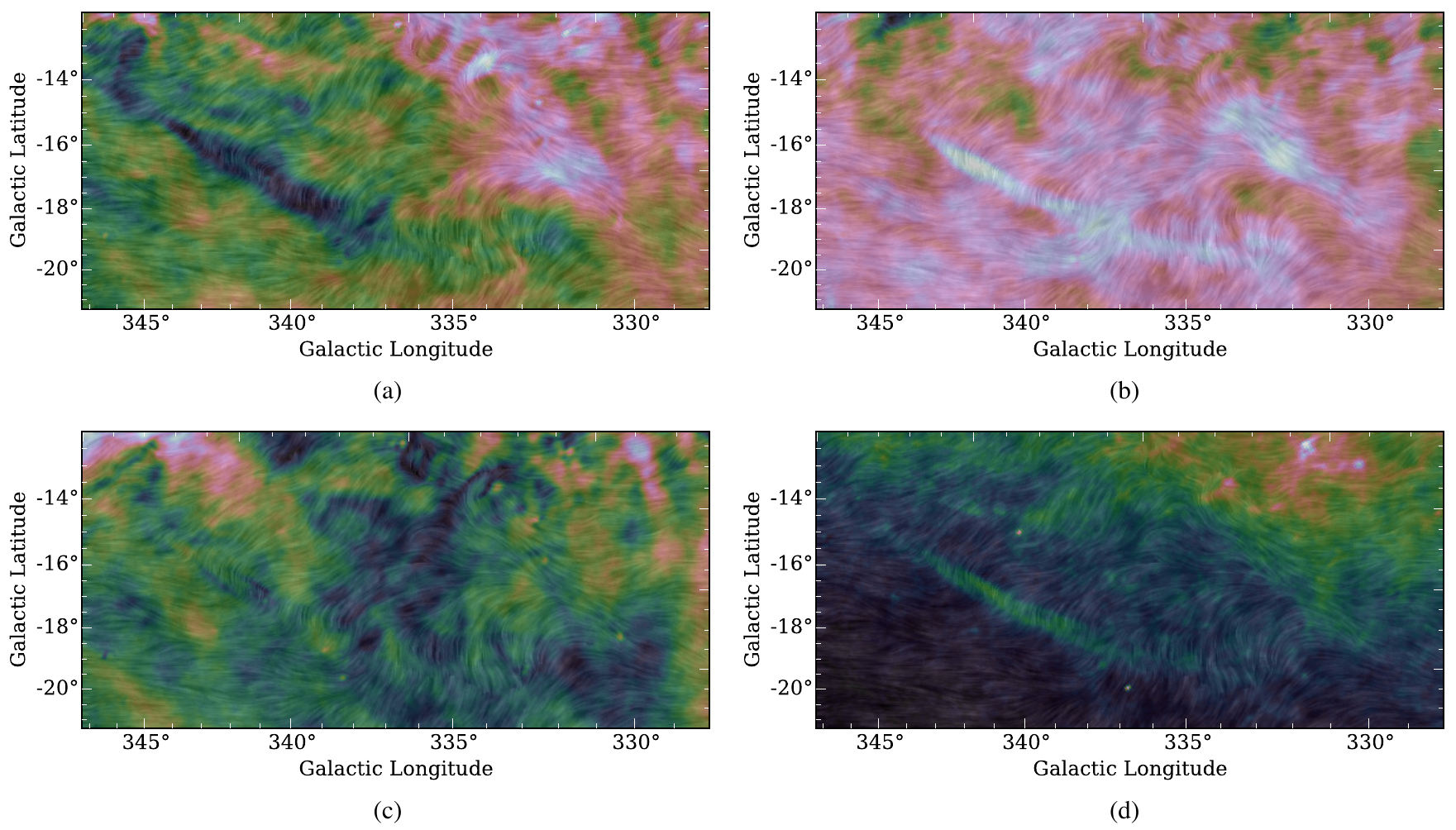}
\caption{(a) and (b) show respectively the $E$- and $B$-mode maps, (c) shows the amplitude of the linear polarisation vector, $|\bmath{P}|$, and (d) shows the H$\alpha$ map. All maps are overlaid with the orientation of the magnetic field using the LIC technique. No colour-bar is displayed on these figures because the drapery structure alters the colour scale.}
\label{fig:EB_lic}
\end{figure*}

The scalar $E$ and pseudo-scalar $B$ maps without applying the differential operators $\eth$ and $\overline{\eth}$ are sensitive to an intervening smooth Faraday screen located along the line of sight; a uniform rotation of the polarisation by an angle $\theta$ induces a fraction $\sin 2\theta$ of the $E$-mode to be converted into $B$-mode \citep{2009PhRvL.102k1302K,2014MNRAS.438.2508P}. Nevertheless, sharp charges in the magnetic field or the free-electron density causing an alignment of the polarisation vector over a filamentary feature would produce an $E$-mode pattern that should also be correlated with the gradient of the polarisation. Figure \ref{fig:gradPvsEB} (a) shows the polarisation gradient of the same region as in Fig. \ref{fig:gradPvstildeEB} (a), but at a larger angular scale of $\sim 161$ arcmin. The feature described by \citet{2014A&A...566A...5I} still appears as a double-jump at this scale. At its original resolution, the distorted double-jump feature may be altered by noise fluctuations. Another double-jump also appears on the right hand side of the subregion. Figures \ref{fig:gradPvsEB} (b) and (c) present, respectively, the $E$- and $B$-mode decomposition without the differential operators $\eth$ and $\overline{\eth}$, i.e. the spherical harmonic synthesis of $a_{E,\ell m}$ and $a_{B,\ell m}$ following equation (\ref{eq:EB_sphe_harm}). All maps in Fig. \ref{fig:gradPvsEB} are overlaid with the $|\nabla\bmath{P}|$ maxima chains at an angular scale of $\sim 161$ arcmin. The maxima chains\footnote{Following the WTMM method \citep{2000EPJB...15..567A}, wavelet transform modulus maxima are positions where $|\nabla\bmath{P}|$ is locally maximum in the direction of arg($|\nabla\bmath{P}|$). Naturally maxima pixels lie on connected ``maxima chains''.} associated with the two double-jump features define well the contours of two positive filament-like features in the $B$-mode map. Using the line integral convolution (LIC) technique \citep{Cabral:1993:IVF:166117.166151}, Figures \ref{fig:EB_lic} (a) and (b) show respectively the $E$- and $B$-mode maps overlaid with the orientation of the magnetic field\footnote{The orientation of the magnetic field is defined as $\chi=\theta+\pi/2$, where $\theta$ is computed using the two-argument arctan function and the IAU convention (see footnote \ref{footnote:atan2}). This definition is only true if the synchrotron emission is not affected by Faraday rotation along the line of sight.}. It can be seen in Fig. \ref{fig:EB_lic} (b) that the two positive filament-like features are indeed located in regions where the direction of the magnetic field changes abruptly to be generally oriented $45^{\circ}$ to these features. On the other hand, the elongated negative region on the left hand side of Fig. \ref{fig:EB_lic} (a) is broader and not as defined as the $B$-mode filament but the vectors orientation is generally perpendicular to the feature. Figures \ref{fig:EB_lic} (c) and (d) show respectively the amplitude of the linear polarisation vector, $|\bmath{P}|$, and the H$\alpha$ emission map of \citet{2003ApJS..146..407F} for the same subregion. Depolarisation canals are present on the edges of the left hand side filament in the $|\bmath{P}|$ map, where changes of the magnetic field orientation are the most abrupt. As noticed by \citet{2014A&A...566A...5I}, the same filament is well correlated with H$\alpha$ emission.

The $E$- and $B$-mode decomposition applied on the polarised synchrotron emission does not carry the same physical meaning than when it is applied for the CMB analysis. Nevertheless, the spin-2 decomposition of the polarised synchrotron emission remains a very interesting strategy to highlight coherent features where a particular alignment of the magnetic field lines occurs. It is demonstrated in section \ref{sec:pow_spec} that instead of applying the differential operators $\eth$ and $\overline{\eth}$, one can calculate the multiscale gradient of the $E$- and $B$-mode maps, i.e. $|\nabla \tilde{E}(l,\bmath{x})|$ and $|\nabla \tilde{B}(l,\bmath{x})|$ (defined in equations (\ref{eq:EB_amplitude})). These quantities are also rotationally invariant in the spatial domain and are sensitive to fluctuations at larger angular scales than the original telescope resolution.

\section{Polarisation gradient of the southern hemisphere}\label{sec:results}

This section shows results of the multiscale gradient of $\bmath{P}$ analysis for the entire S-PASS survey and for some subregions of interest. For these subregions, a comparison is done with structures seen in $E$- and $B$-mode maps.

\subsection{Galaxy-scale and Halo}\label{sec:galaxy_halo}

\begin{figure*}
\centering
\includegraphics[]{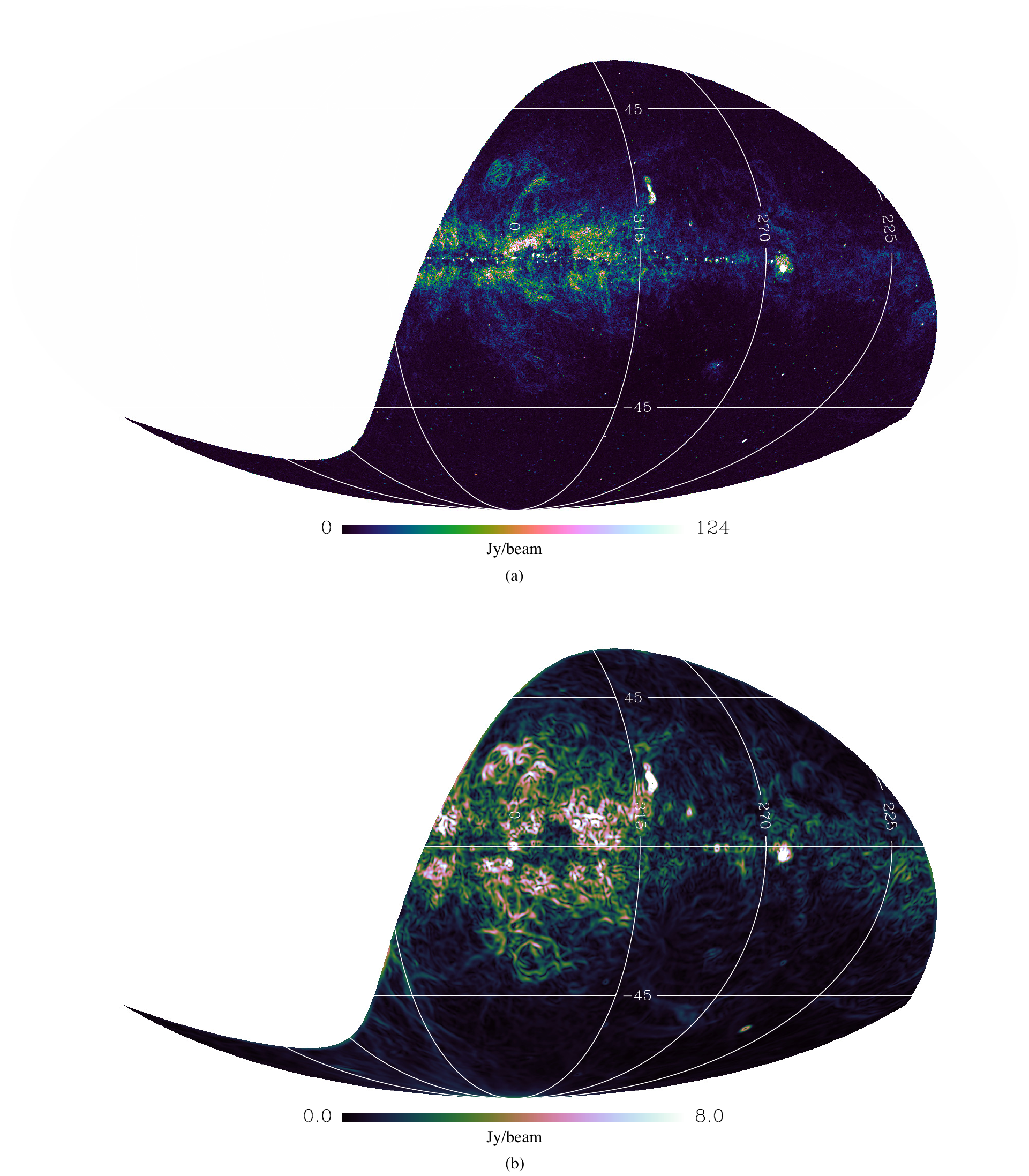}
\caption{The spatial gradient of the polarised synchrotron emission at 2.3 GHz for the southern hemisphere at the original resolution of the S-PASS survey (top) and at scale of approximately 3 degrees (bottom). Maps are the Mollweide projections of the \texttt{HEALPix} pixelisation maps.}
\label{fig:mollweide_gradpA}
\end{figure*}

\begin{figure*}
\centering
\includegraphics[]{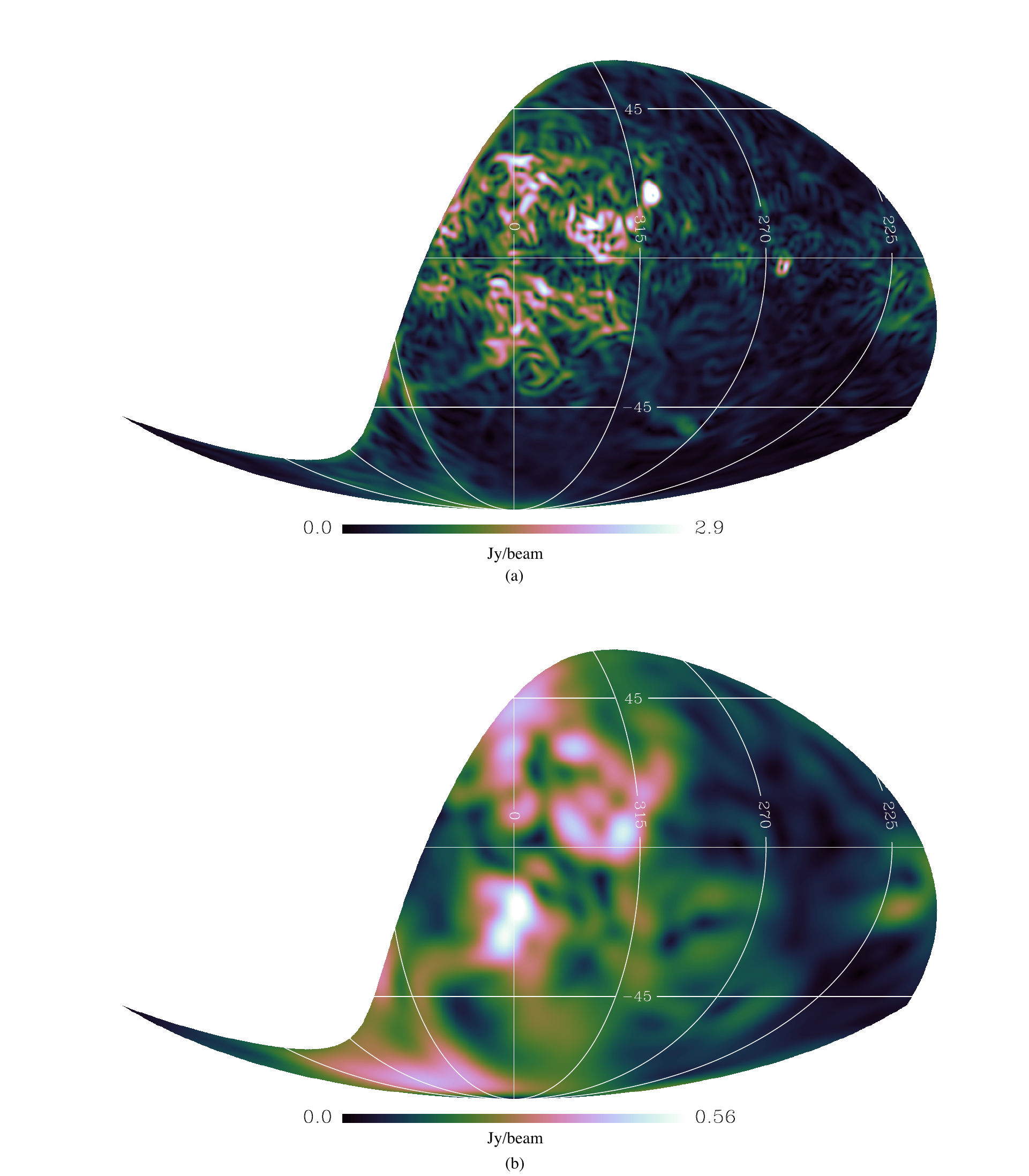}
\caption{Same as Figure \ref{fig:mollweide_gradpA} at scales of approximately 8 (top) and 33 (bottom) degrees.}
\label{fig:mollweide_gradpB}
\end{figure*}

\begin{figure*}
\centering
\includegraphics[]{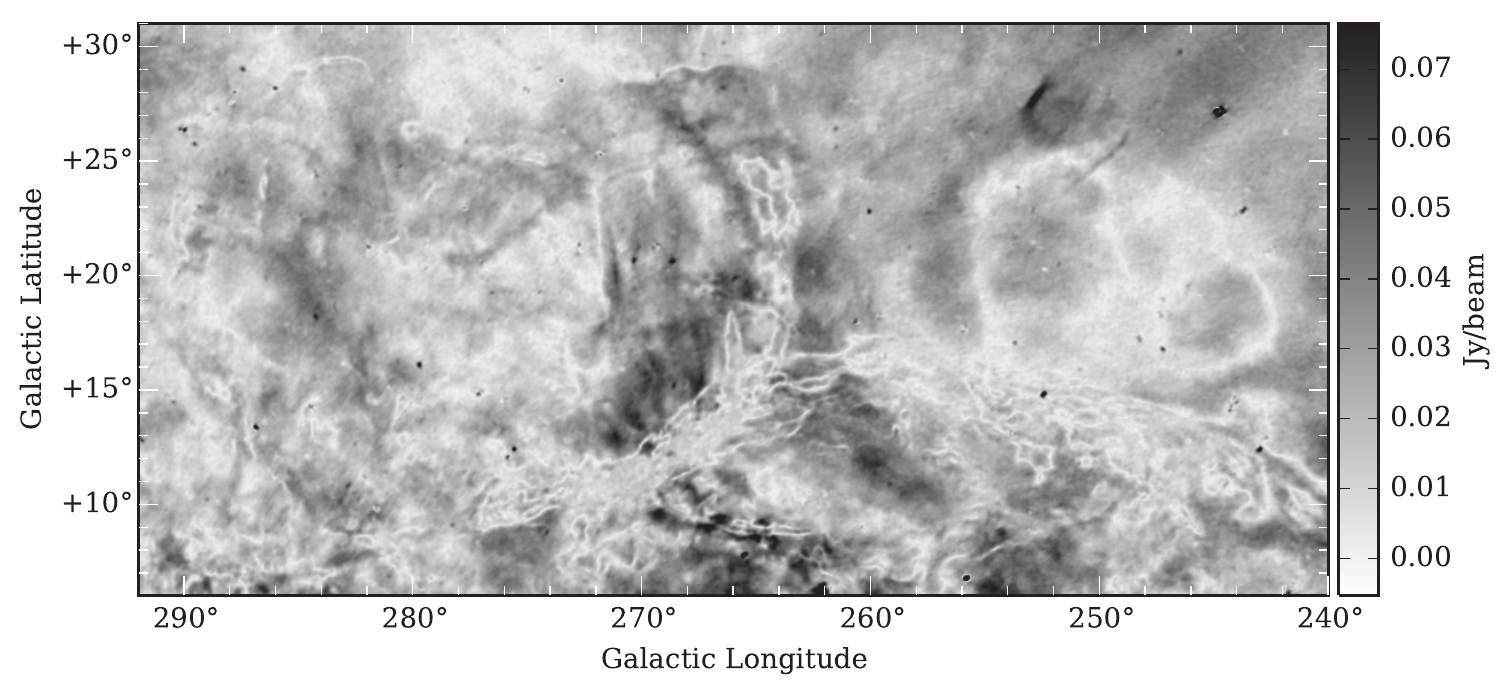}
\caption{$|\bmath{P}|$ of a subregion of the S-PASS data including the old SNRs Antlia \citep{2002ApJ...576L..41M} (the hearth shape on the left hand side of the figure) and a part of the Gum Nebula \citep{2015ApJ...804...22P} (the arc at the bottom half on the right hand side of the figure) where many depolarisation canals are present.}
\label{fig:depolarisation}
\end{figure*}

Figures \ref{fig:mollweide_gradpA} and \ref{fig:mollweide_gradpB} show the spatial gradient of polarised synchrotron emission at 2.3 GHz for the southern hemisphere at the original resolution of the S-PASS survey, and at scales of approximately 3, 8 and 33 degrees. These southern hemisphere all-sky views show the $|\nabla \bmath{P}|$ maps instead of $|\nabla \bmath{P}|/|\bmath{P}|$ as shown by \citet{2014A&A...566A...5I}. This choice is justified by the large number of depolarisation canals in the $|\bmath{P}|$ map. Figure \ref{fig:depolarisation} shows a subregion including the old SNR Antlia \citep{2002ApJ...576L..41M} where many depolarisation canals are present. This effect, caused by beam depolarisation, occurs when the telescope beam is larger than the scale of the RM gradient \citep{2004A&A...421.1011H}. Depolarisation canals are present mainly in highly turbulent regions where the gradient of the polarisation vector is so high that the beam averaging cancels the polarised emission. Consequently, the division of $|\nabla \bmath{P}|$ by the very low polarisation intensity of those regions falsely increases the value of the polarisation gradient. Nevertheless, since depolarisation canals trace sharp changes of the polarisation intensity as a function of the position, these canals are also correlated with structures seen in $|\nabla \bmath{P}|$ at the original resolution but do not show very high intensity filaments as in the $|\nabla \bmath{P}|/|\bmath{P}|$ maps.

The very large column density of the magneto-ionic medium and the dynamical complexity of the Galactic centre also makes the polarised signal in large areas near the Galactic centre highly depolarised \citep{2013Natur.493...66C}. For this reason, polarisation gradients at the original resolution and at 3 and 8 degrees in Fig. \ref{fig:mollweide_gradpA} and \ref{fig:mollweide_gradpB} (a) show depth depolarisation in the Galactic plane near the Galactic centre. Despite this similarity, each scale shows greater differences in the filament network across the sky than might be expected from simply a smoothed version of the previous angular scale. As described in section \ref{sec:multiscale}, the polarisation gradient operation applied on different scales acts as a band-pass filter in the frequency domain and thus reveals fluctuations in the data that are present over a certain range of scales only. The most obvious difference is the amount of structure at high and low Galactic latitudes. This effect can be explained by the fact that some large-scale structures are located closer to us and thus cover a larger angular size in latitude.

\begin{figure}
\centering
\includegraphics[]{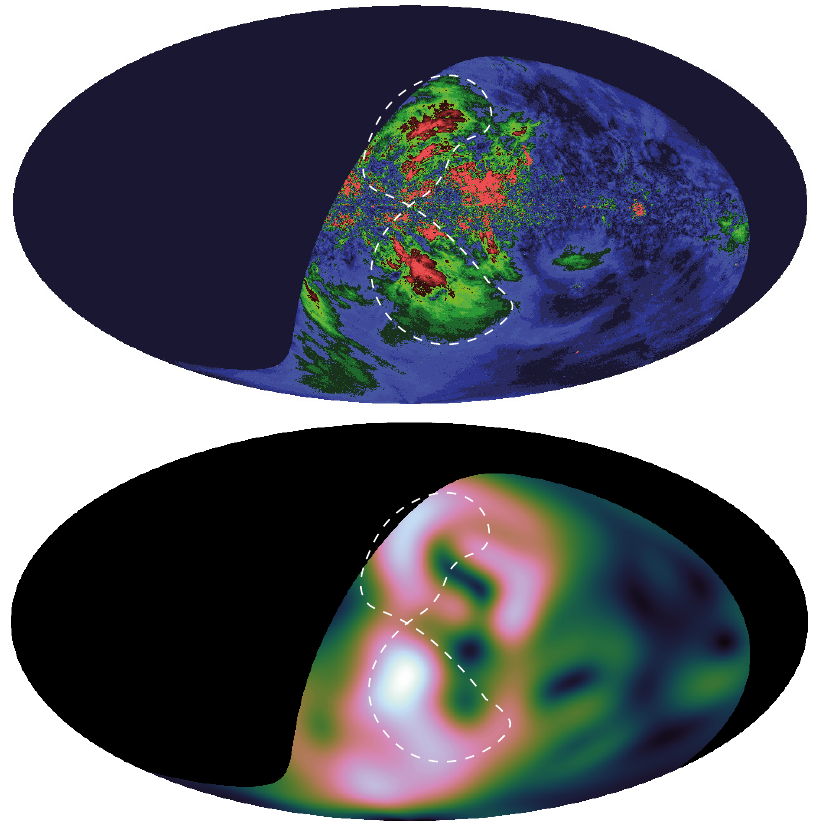}
\caption{Top: Polarisation intensity of S-PASS data using the same colour scale and Aitoff projection as \citet{2013Natur.493...66C}, Fig. 1. Bottom: $|\nabla \bmath{P}|$ at scale of $\sim 66^{\circ}$ on Aitoff projection. The
dashed line delineate the giant radio lobes.}
\label{fig:halo}
\end{figure}

Nevertheless, it can be seen that most of the large-scale structures above a Galactic latitude of $b\gtrsim 10^{\circ}$ are concentrated between Galactic longitudes $45^{\circ} \gtrsim l \gtrsim 315^{\circ}$. At a scale of $\sim 30^{\circ}$ in Fig. \ref{fig:mollweide_gradpB} (b) we found structures that correspond well to the giant magnetised outflows identified by \citet{2013Natur.493...66C}, themselves corresponding closely to the Fermi bubbles revealed in $\gamma$-ray emission \citep{2010ApJ...724.1044S}. As shown in Fig. \ref{fig:halo}, this correlation becomes stronger at larger scales. According to \citet{2013Natur.493...66C}, the large-scale emission associated with these giant magnetised outflows must be located behind the depolarising objects seen towards the centre of the Galactic plane. Those depolarisation features are spatially correlated with the H$\alpha$ emission of objects mostly located in the Sagittarius arm, which implies a minimal distance of 2.5 kpc for the outflow lobes. They concluded that the radio lobes are probably star-formation-driven outflows associated with the molecular gas ring occupying the Galactic centre. The high polarisation fraction and the spectral index value of $-1.0$ to $-1.2$ of the lobes are good indications that these outflows originate from the plane, driven by the intense star-forming molecular gas pushing the ionised gas and the frozen-in magnetic field lines in the Galactic halo. The diffuse emission of the lobes is only visible at scales larger than $\sim 10^{\circ}$, while smaller fluctuations associated with the ridges and the Galactic Centre spur, also discussed by \citet{2013Natur.493...66C}, appear at smaller scales and at the original resolution of the $|\nabla \bmath{P}|$ map. The large-scale features correlated to the giant lobes at the bottom of Fig. \ref{fig:halo} extend outside the previous edges delimiting the polarisation intensity. This extension in the $|\nabla \bmath{P}|$ map can be explained mainly by two factors. The first factor is that the spatial gradient of the linear polarisation vector traces edges of features where changes in the polarisation vector occur. According to that definition, the $|\nabla \bmath{P}|$ map should follow the dashed line of Fig. \ref{fig:halo} and not be enclosed by it. This description fits well with the bottom of the southern lobe. The extension of the southern lobe outside the dashed line is likely to be associated with the New Loop described by \citet{2007ApJ...664..349W} and the VIIb loop described by \citet{2015MNRAS.452..656V}. The second factor is that the gradient of the linear polarisation is sensitive to spatial variations of the polarisation intensity and to spatial variations of the polarisation angle. This property of the $|\nabla \bmath{P}|$ map could explain the continuation, which seems to close the two loops, for both the northern and the southern lobes.

\subsection{\HII regions, SNRs and small-scale features}\label{sec:thermal_magnetic}

This section focus on peculiar regions, notably some reported by \citet{2014A&A...566A...5I} in section 3.3. These are structures induced by electron density and/or magnetic field strength fluctuations associated with nearby \HII regions, nearby and old SNRs,  or simply singular  regions because of their intensity or peculiar shape.

\begin{figure*}
\centering
\includegraphics[]{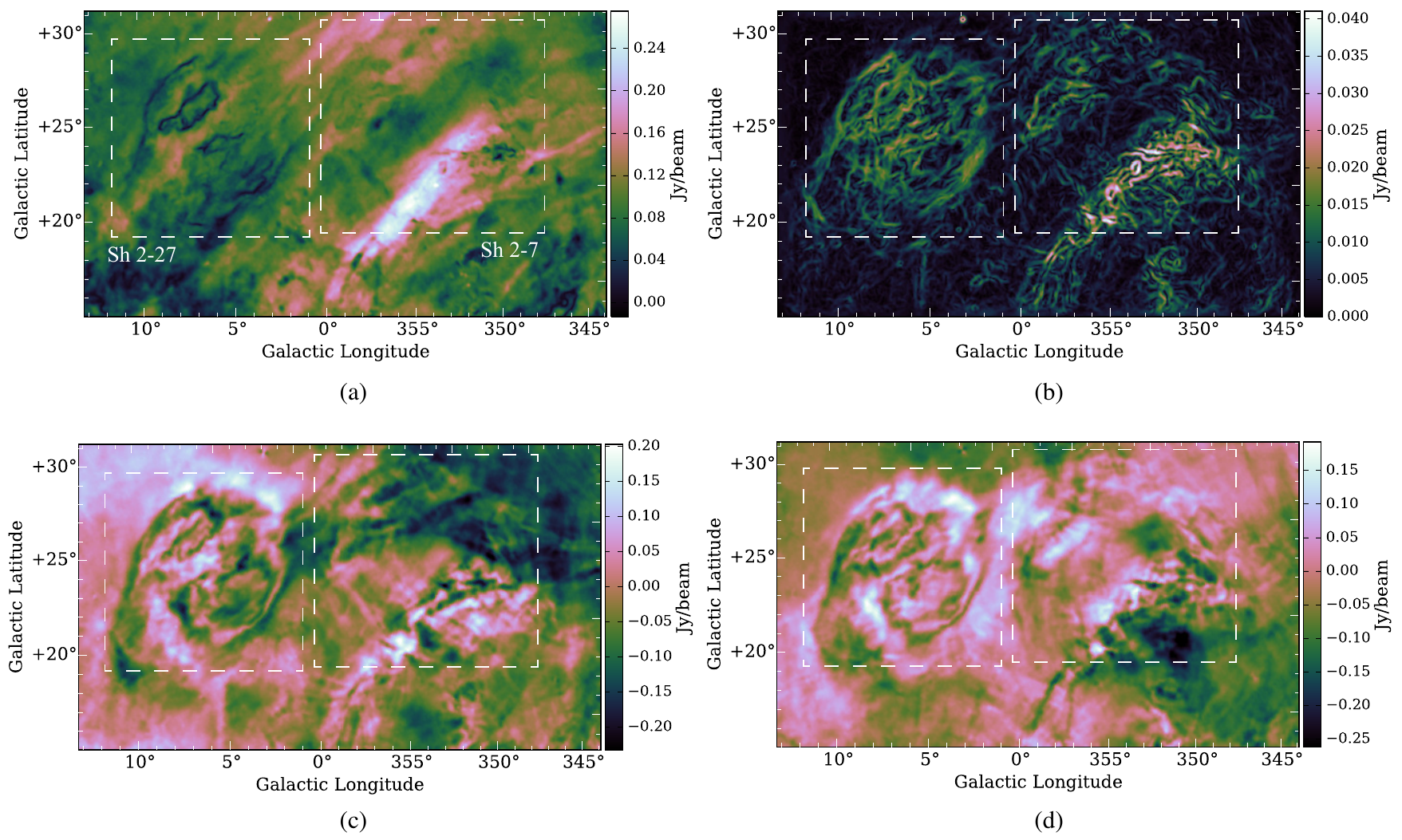}
\caption{A subregion of the S-PASS survey including the \HII regions Sh 2-7 and Sh 2-27. (a) shows the polarised intensity, (b) shows the $|\nabla \bmath{P}|$ map at its original resolution, and the $E$- and $B$-mode maps associated with the same region are shown in (c) and (d).}
\label{fig:Sh2-7_Sh2-27}
\end{figure*}

\citet{2014A&A...566A...5I} discussed two nearby \HII regions, Sh 2-7 and Sh 2-27, which have a clear spatial correlation with H$\alpha$ emission. Both of these \HII regions in Fig. \ref{fig:Sh2-7_Sh2-27} (a) show depolarisation canals in the $|\bmath{P}|$ map, thus a direct comparison of their normalised polarisation gradient features, $|\nabla \bmath{P}|/|\bmath{P}|$, with MHD turbulence simulations representing only foreground Faraday fluctuations should be interpreted carefully. Despite the fact that features in the southern part of Sh 2-7 are affected by the overlap of the Galactic Centre spur, \HII regions Sh 2-7 and Sh 2-27 appear well defined in the $|\nabla \bmath{P}|$ map (Fig. \ref{fig:Sh2-7_Sh2-27} (b)) compared to the $|\bmath{P}|$ map. This confirms that fluctuations seen in $|\nabla \bmath{P}|$ are foreground Faraday rotation fluctuations produced by the \HII regions. At larger scales, Sh 2-7 is dominated by features of the Galactic Centre spur and diffuse emission. Figures \ref{fig:Sh2-7_Sh2-27} (c) and (d) show respectively the $E$- and $B$-mode maps of the same subregion. Contrary to the polarisation intensity map shown in Fig. \ref{fig:Sh2-7_Sh2-27} (a), the $E$- and $B$-mode maps are sensitive to both variations of the polarisation vector, the polarisation intensity and its position angle. For this reason, fluctuations caused by Faraday rotation in  \HII regions Sh 2-7 and Sh 2-27 appear on these maps.

\begin{figure*}
\centering
\includegraphics[scale=1.0]{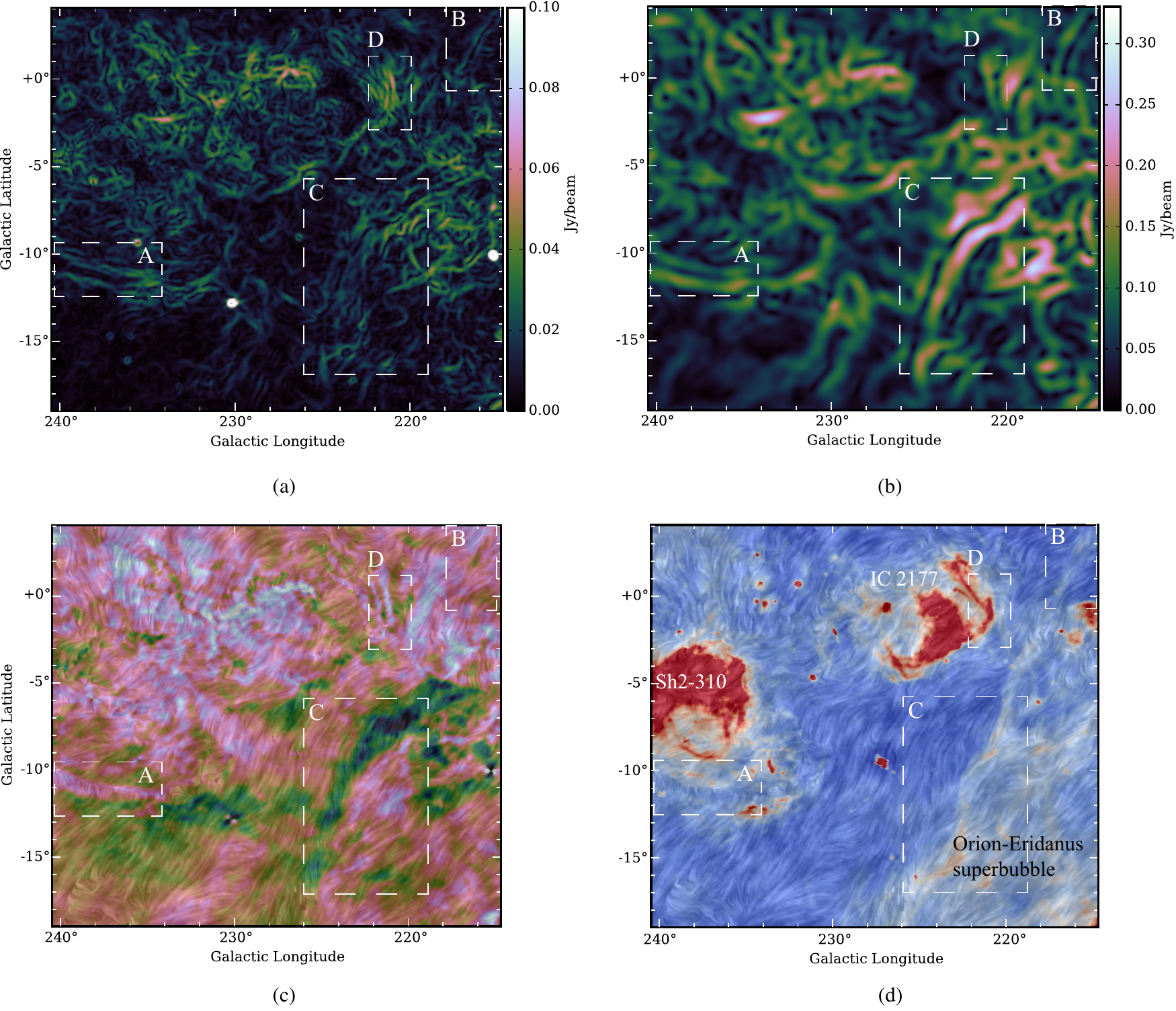}
\caption{Anti-centre Galactic field including the \HII region Sh2-310, the nebula IC 2177 and a part of the Orion-Eridanus superbubble. The panel (a) shows the $|\nabla \bmath{P}|$ map at the original resolution, panel (b) shows the $|\nabla \bmath{P}|$ map at $\sim 136$ arcmin, panel (c) shows the $B$-mode map overlaid with the magnetic field line direction in drapery texture and panel (d) shows the $H\alpha$ overlaid with the same drapery texture. White dashed boxes indicate areas where $|\nabla \bmath{P}|$ and $B$-mode features are correlated. No colour-bar is displayed on Figures (c) and (d) because the drapery structure alters the colour scale.\label{fig:Eridanus}}
\end{figure*}

In the direction of the Galactic anti-centre, a complex region located between the Gum nebula and the Orion-Eridanus superbubble shows several correlated features between $|\nabla \bmath{P}|$, the $B$-type polarisation and $H\alpha$. Figure \ref{fig:Eridanus} (a) and (b) show respectively the $|\nabla \bmath{P}|$ map at its original resolution and at $\sim 136$ arcmin. At the original resolution, many double-jump morphology features are visible. Some of them are still visible at larger scale, as the feature in box B in Fig. \ref{fig:Eridanus} (b) and others become more definite at larger scales, as the features in boxes A and C. On the other hand, the double-jump feature with the higher intensity in Fig. \ref{fig:Eridanus} (a) (box D), which marks the edge of the nebula IC 2177, seems to disappear in Fig. \ref{fig:Eridanus} (b). The double-jump B is well correlated with a filament in $H\alpha$ marking the upper-edge of the Orion-Eridanus superbubble (see Fig. \ref{fig:Eridanus} (d)). The feature A is located between two filamentary structures in $H\alpha$ likely associated with the \HII region Sh2-310. The latter also appears as a positive filament in the $B$-mode map in Fig. \ref{fig:Eridanus} (c). Similarly, feature C, which appears as a negative filament in the $B$-mode map, is also located at the edge of a $H\alpha$ filamentary structure. The positive or negative value of a feature in the $B$-mode map depends on the relative direction of the $45^{\circ}$ alignment of the polarisation vector with changes in the polarisation intensity. For this particular region, this morphology might be caused by the compression of magnetic field lines generally aligned with the expanding bubble.

\section{Influence on Power Spectra analysis}\label{sec:pow_spec}

\subsection{$|\nabla \bmath{P}|$ vs $|\nabla \bmath{E}|$ and $|\nabla \bmath{B}|$ power spectra}\label{sec:pow_spec_PEB}

The analysis of complex regions such as the one presented in Fig. \ref{fig:Eridanus} shows that polarisation patterns creating $B$-mode filament-like features are correlated with double-jump features in $|\nabla \bmath{P}|$ maps and that they can appear at different angular scales. Since \citet{2015MNRAS.451..372R} have shown that particularly intense fluctuations, such as double-jumps, can influence the power spectrum analysis of a region, it is justified to test if the peculiar features found in Fig. \ref{fig:Eridanus} influence the power spectra of $E$ and $B$ maps.

We introduce a new quantity $|\nabla EB|$ that applies the concept of gradient to $E$- and $B$-modes. Inspired by the definitions in equations (\ref{eq:scaled_gradient_P}) and (\ref{eq:Stokes_amplitude}), it is defined in the wavelet space as:

\begin{equation}
|\nabla \tilde{E}\tilde{B}(l,\bmath{x})| = \sqrt{ |\nabla \tilde{E}(l,\bmath{x})|^2 + |\nabla \tilde{B}(l,\bmath{x})|^2},
\label{eq:scaled_gradient_EB}
\end{equation}

\begin{equation}
\begin{split}
|\nabla \tilde{E}(l,\bmath{x})| & = \sqrt{ |\tilde{E}_{1}(l,\bmath{x})|^2 + |\tilde{E}_{2}(l,\bmath{x})|^2},\\
\\
|\nabla \tilde{B}(l,\bmath{x})| & = \sqrt{ |\tilde{B}_{1}(l,\bmath{x})|^2 + |\tilde{B}_{2}(l,\bmath{x})|^2},\\
\end{split}
\label{eq:EB_amplitude}
\end{equation}

\noindent where indices 1 and 2 refer to the two directional wavelet transforms of equations (\ref{eq:DoG_transforms}) and (\ref{eq:Gauss_derivative}). Because using a wavelet with more vanishing moments can represent more complex signals \citep{2007hefm.book.....T}, the third order wavelet, i.e. $m=3$ in equation (\ref{eq:Gauss_derivative}), is used to calculate the wavelet coefficients involved in the wavelet power spectrum, which is defined as 

\begin{equation}
S_X(l) = \langle |\nabla \tilde{X}(l,\bmath{x})|^2 \rangle_{\bmath{x}},
\label{eq:wav_pow_spec}
\end{equation}

\noindent where $\tilde{X}$ can be $\tilde{\bmath{P}}$, $\tilde{E}$, $\tilde{B}$ or $\tilde{E}\tilde{B}$ and $\langle\rangle_{\bmath{x}}$ represents the average operation over the two-dimensional coordinate $\bmath{x}$.

\begin{figure*}
\centering
\includegraphics[]{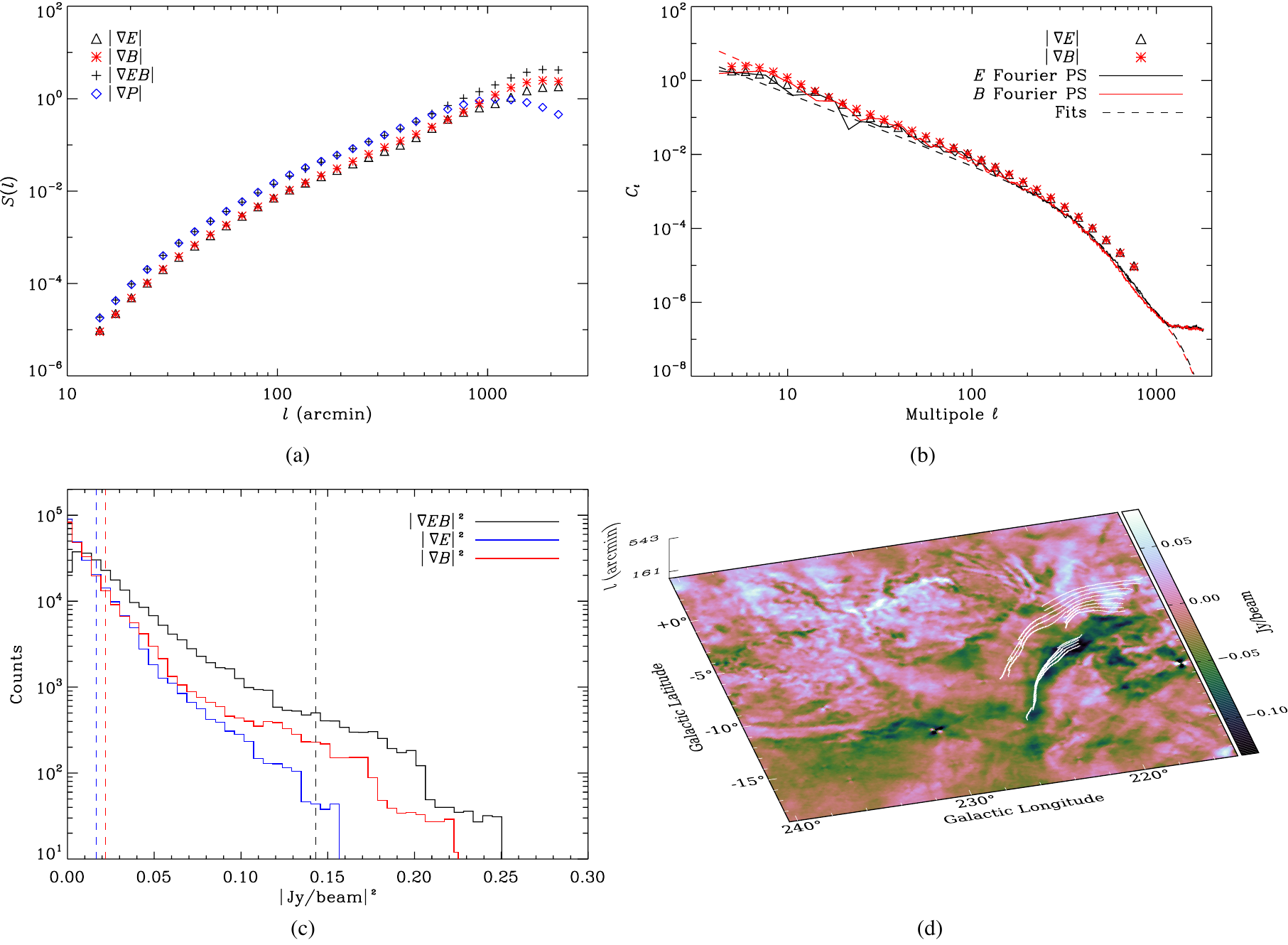}
\caption{(a)The wavelet power spectrum of $|\nabla \bmath{P}|$ compared with wavelet power spectra of $|\nabla E|$, $|\nabla B|$ and $|\nabla EB|$ for the Orion-Eridanus region. (b) Wavelet power spectra of $|\nabla E|$ and $|\nabla B|$ compared with the Fourier power spectra of $E$- and $B$-mode maps. (c) shows the histogram of the power distribution for $l\simeq271$ arcmin. The blue and red dashed lines represent, respectively, the mean value for $|\nabla E|^2$ and $|\nabla B|^2$ distributions and the black dashed line represents the chosen threshold delimiting the power excess. (d) is the $B$-mode map of the Orion-Eridanus region overlaid with the $B$-mode maxima chains excess for scales $161 \lesssim l \lesssim 543$ arcmin. \label{fig:pow_spec}}
\end{figure*}

Figure \ref{fig:pow_spec} (a) shows the wavelet power spectrum of $|\nabla \bmath{P}|$ compared with wavelet power spectra of $|\nabla E|$, $|\nabla B|$ and $|\nabla EB|$ of the Orion-Eridanus region. Since the scalar $E$ and pseudo-scalar $B$ are tracing together all fluctuations of the polarisation vector $\bmath{P}$, the wavelet power spectrum of $|\nabla \bmath{P}|$ should, in the small-scale limit, be equal to the wavelet power spectrum of $|\nabla EB|$ (see Appendix \ref{sec:appendix} for the full derivation). We can see in Fig. \ref{fig:pow_spec} (a) that this equality is true for scales $l\lesssim 600$ arcmin. The inequality between $S_P$ and $S_{EB}$ at larger scales arises from the fact that the $E$- and $B$-mode decomposition is non-local. Consequently, polarisation patterns contributing to the $E$- and $B$-mode features at large scales can be located outside the field. We must recall that the $E$- and $B$-mode decomposition is done using all data projected on a sphere. According to Fig. \ref{fig:pow_spec} (a), there are also small deviations between $S_E$ and $S_{B}$ for $160 \lesssim l \lesssim 550$ arcmin with a mean variance ratio $S_{B}/S_{E}=1.14$. The smaller excess of $B$-mode for $25 \lesssim l \lesssim 70$ arcmin has a mean variance ratio of $S_{B}/S_{E}\approx1.04$. The larger deviation between $E$ and $B$ spectra at larger scales might be caused by the same effect as the deviation between $S_P$ and $S_{EB}$, i.e. the non-local property of the $E$- and $B$-mode decomposition. For $l \gtrsim 900$ arcmin the mean variance ration is $S_{B}/S_{E}\approx1.43$.

Figure \ref{fig:pow_spec} (b) shows the $|\nabla E|$ and $|\nabla B|$ wavelet power spectra overlaid with Fourier power spectra of the same regions. Fourier power spectra are defined as:

\begin{equation}
C^{\rm{F}}(k) = \langle |\hat{f}(k)|^2 \rangle_{\theta},
\label{eq:fourier_pow_spec}
\end{equation}

\noindent where $\hat{f}(k)$ is the Fourier transform of the image, i.e. the $E$- or $B$-mode map, and $\langle \rangle_{\theta}$ is the average operation over the azimuthal direction $\theta$. In Fig. \ref{fig:pow_spec} (b), the wavenumber $k$ is converted to the angular scale $l$ according to the relation $k=1/l$ (see section \ref{sec:multiscale}) and then converted to multipoles $\ell$ following the relation $l=180^{\circ}/\sqrt{\ell(\ell+1)}$. In the small-scale approximation, this calculation of the $E$- and $B$-mode power spectra is equivalent to equation (\ref{eq:Sphe_pow_spec}). Even if asymmetric wavelets are not designed to reproduce the Fourier power spectrum \citep{2005CG.....31..846K}, both spectra are generally similar and deviations between $E$- and $B$-type fluctuations noticed in wavelet's power spectra are also visible in the Fourier power spectra. However, the smooth shape of the wavelet function produces a smoother spectrum where significant deviations can be identified more easily.

As noticed by \citet{2015MNRAS.451..372R}, the wavelet power spectra do not sample properly the drop of power caused by the telescope beam. Thus, in order to measure power laws corrected for the effective beam, Fourier power spectra have been fitted with a similar relation to that of \citet{2010MNRAS.405.1670C} :

\begin{equation}
C(\ell) = \left[C_{200}\left(\frac{\ell}{200}\right)^{\beta}W_{\ell}^{B} + N \right] W_{\ell}^{C},
\label{eq:fit}
\end{equation}

\noindent where $C_{200}$ is the spectrum at $\ell=200$, $W_{\ell}^{B}$ is the telescope beam window function with FWHM = 8.9 arcmin, and $W_{\ell}^{C}$ a Gaussian window function of FWHM = 6 arcmin modelling the convolution of the final maps (see section \ref{sec:observation}). The noise level $N$ is not uniform across the S-PASS survey, however it has been modelled as a constant for the analysis of subregions. Both Gaussian window functions, $W_{\ell}^{B}$ and $W_{\ell}^{C}$, are fixed for the fit. Fitting values are summarised in Table \ref{tab:fit}.

\begin{table*}
\caption{Fitting values for the Fourier power spectra between multipoles $\ell=20$ and 1150.}
\label{tab:fit}
\begin{tabular}{lcccc}
\hline
Field			&	$C_{200}^{E}$						&	$\beta^{E}$		&	$C_{200}^{B}$						&	$\beta^{B}$	\\
			&	($\times10^{-3}$ $|$Jy/beam$|^2$)		&					&	($\times10^{-3}$ $|$Jy/beam$|^2$)		&				\\
\hline
Orion-Eridanus	&	$1.4 \pm 0.1$						&	$-1.9 \pm 0.1$		&	$1.4 \pm 0.1$						&	$-2.2 \pm 0.1$\\
G353-34		&	$1.2\pm 0.1$						&	$-2.8 \pm 0.1$		&	$1.1 \pm 0.1$						&	$-3.0 \pm 0.1$\\
\hline
\end{tabular}
\end{table*}

\subsection{Locating $E$- and $B$-mode asymmetries}\label{sec:asymmetries}

The main advantage of using wavelet coefficients instead of Fourier coefficients to evaluate the power spectrum of an image is that wavelet coefficients' spatial distribution can be analysed as a function of scale. This analysis can be used to locate specific features responsible for the $E$- or $B$-type polarisation excess in the power spectrum. As shown in Fig. \ref{fig:pow_spec}, $|\nabla EB|$ shows no particular break in its power spectrum. Since it represents the total power for that region, depending on the polarisation patterns present at a given scale, we can see that the total power is sometimes distributed evenly between $|\nabla E|$ and $|\nabla B|$ power spectra and sometimes it is distributed unequally. Figure \ref{fig:pow_spec} (c) shows the histogram of the squared coefficient distribution using the first ordered wavelet, i.e. $m=1$, for the angular scale $l\simeq271$ arcmin, which represents approximately the middle of the excess in the $B$-mode power spectrum. Both fields are non-Gaussian, however according to equations (\ref{eq:scaled_gradient_EB}) and (\ref{eq:EB_amplitude}), the distributions should be comparable to non-central chi-squared distributions with two degrees of freedom for $|\nabla E|^2$ and $|\nabla B|^2$, and four degrees of freedom for $|\nabla EB|^2$ with a non-unitary variance. The blue and red dashed lines represent, respectively, the mean values for $|\nabla E|^2$ and $|\nabla B|^2$ distributions and the black dashed line represent the chosen threshold delimiting the power excess. The threshold has been set at $4\sigma$ over the mean value of $|\nabla EB|^2$ divided by 2.

In order to display features associated with the excess for a range of scales, only the maximal pixels along each maxima chain are considered. If a maximal pixel value is above the threshold, then the entire associated maxima chain is displayed. Figure \ref{fig:pow_spec} (d) displays the $B$-mode map of the Orion-Eridanus region, overlaid with the $B$-mode maxima chains excess for scales $161 \lesssim l \lesssim 543$ arcmin. For this range of scales, only one region contributes significantly to the excess in the $B$-type polarisation. This region is correlated with the Orion-Eridanus superbubble edge where a filament-like feature in $B$-mode is also present.

\begin{figure*}
\centering
\includegraphics[]{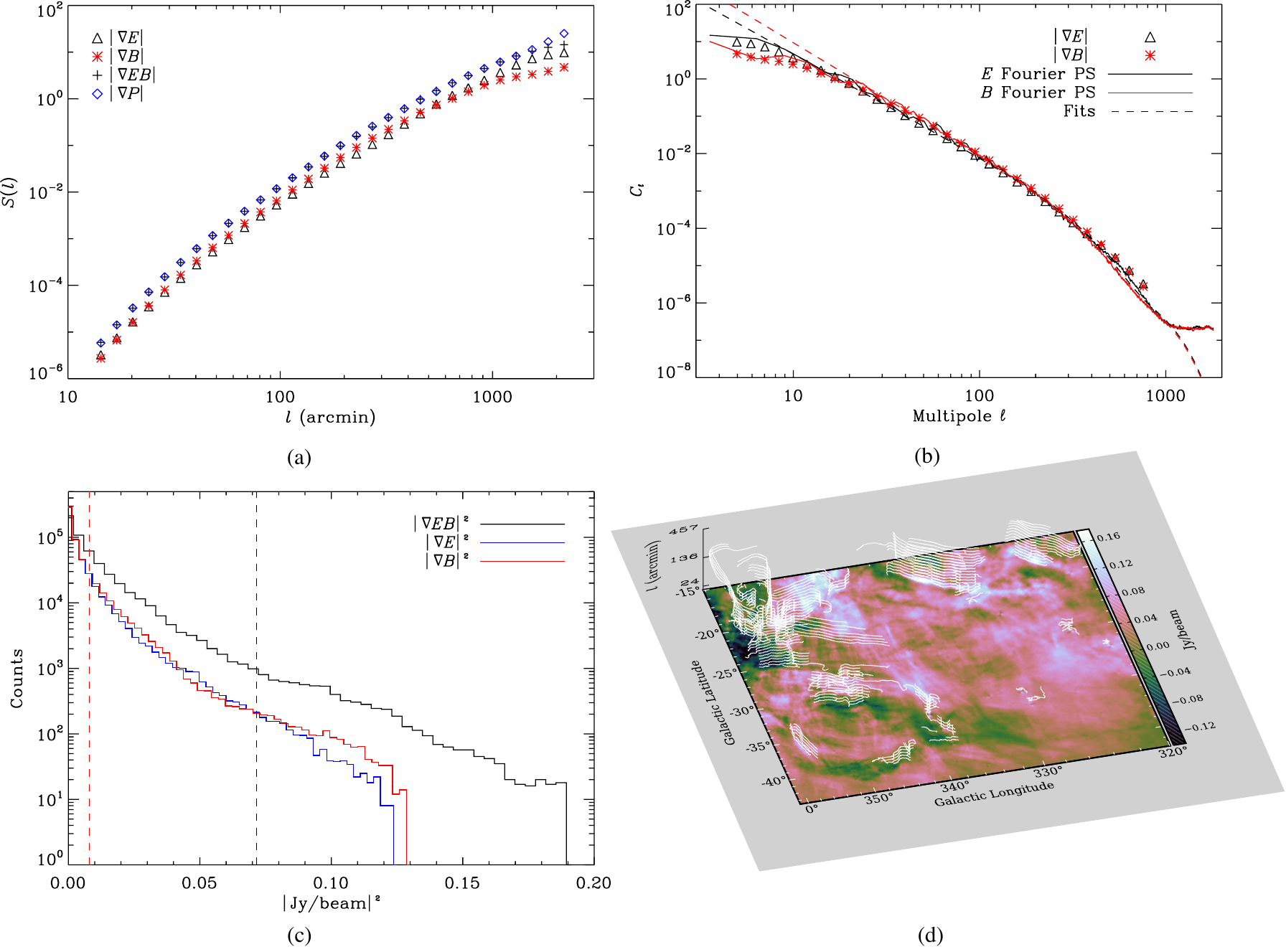}
\caption{(a)The wavelet power spectrum of $|\nabla \bmath{P}|$ compared with wavelet power spectra of $|\nabla E|$, $|\nabla B|$ and $|\nabla EB|$ for the G353-34 region. (b) shows wavelet power spectra of $|\nabla E|$ and $|\nabla B|$ compared with the Fourier power spectra of $E$- and $B$-mode maps.
 (c) shows the histogram of the power distribution for scale $l\sim136$ arcmin. The blue and red dashed lines represent respectively the mean value for $|\nabla E|^2$ and $|\nabla B|^2$ distributions and the black dashed line represents the chosen threshold delimiting the power excess. (d) is the $B$-mode map of the G353-34 region over laid with the $B$-mode maxima chains excess for scales $24 \lesssim l \lesssim 457$ arcmin.\label{fig:excessB}}
\end{figure*}

The same analysis has also been performed on another region located at higher Galactic latitude, including the field shown in Fig. \ref{fig:EB_lic} and the G353-34 shell structure \citep{2008A&A...484..733T}. Results are shown in Fig. \ref{fig:excessB}. The power spectrum shows significant asymmetries between $E$- and $B$-mode scales $25 \lesssim l \lesssim 500$ arcmin. The variance ratio in this range of scales is $S_{B}/S_{E} = 1.19$. As shown in Table \ref{tab:fit}, both $E$ and $B$ Fourier power spectra have a steeper slope compared to the Orion-Eridanus region. The histogram of Fig. \ref{fig:excessB} (c) shows the distribution of the wavelet coefficients for $l\simeq136$ arcmin, which represents approximately the middle of the excess in the $B$-mode power spectrum. Figure \ref{fig:excessB} (d) shows the $B$-mode map overlaid with the maxima chains excess. The excess at the top right of the map is correlated with the $B$-mode filament-like counterparts discussed in section \ref{sec:gardPvsSpin2} and displayed in Fig. \ref{fig:EB_lic} (a). It is interesting to note the coherency of these excesses over a large range of spatial scales. The second coherent excess is located around the G353-34 shell structure. Finally, a significant amount of the excess is located above the G353-34 shell. Parts of these structures are also visible in the $|\bmath{P}|$ map. The feature in the $|\bmath{P}|$ map is more extended than in $E$- and $B$-mode maps and had been identified as a possible southern extension to the Loop I, also called the North Polar Spur \citep{2015MNRAS.452..656V} . However, it is still debated and there is no proof yet that it is a southern extension of Loop I. The feature has also been identified by \citet{2013Natur.493...66C} as the Southern Ridge. The structure is well known for its coherent magnetic field lines along the filament, which might explain why it produces $E$- and $B$-mode patterns.

\section{Discussion}\label{sec:discussion}

Similarly to the previous work introducing the multiscale analysis of the gradient of $\bmath{P}$ \citep{2015MNRAS.451..372R}, the polarisation fluctuation analysis of the entire southern hemisphere has revealed different networks of filaments as a function of angular scale. The gradient of $\bmath{P}$ analysis at small and intermediate scales reveals complex morphologies through the entire Galactic plane and also at high Galactic latitude. The fact that some uncertain morphological structures at the original resolution, e.g. single or double morphology features, become clearly associated with $B$-mode filament-like structures at larger angular scales implies that a direct analysis of $|\nabla\bmath{P}|$ morphological structures applied at a unique resolution must be considered carefully. Morphological interpretation of some features can be biased by the presence of noise at smaller scales. Moreover, by applying the gradient at the original resolution only, large structures associated with large-scale and distant MHD processes are filtered out and are not considered in the analysis.

\begin{figure*}
\centering
\includegraphics[]{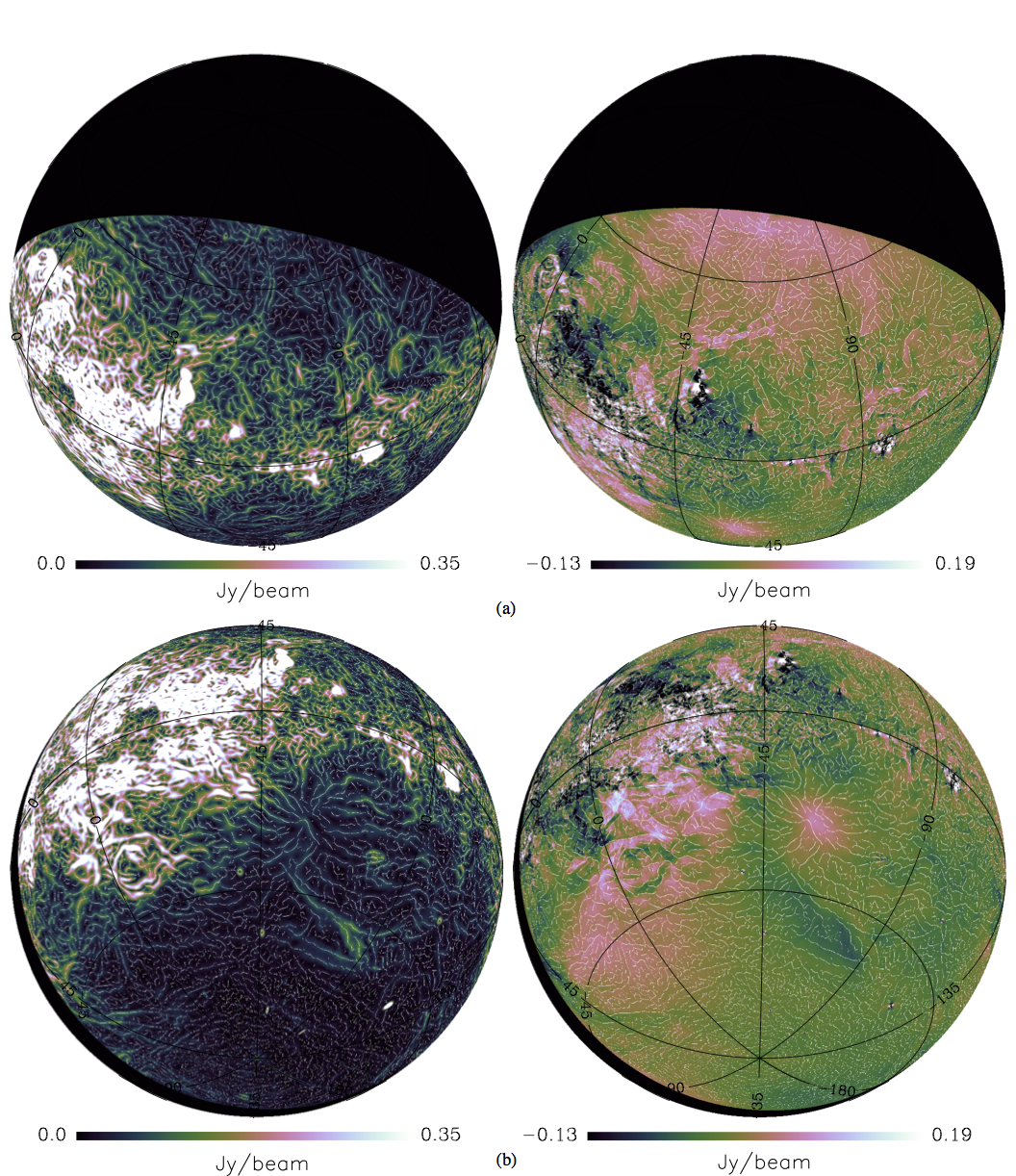}
\caption{Comparison between the $|\nabla \bmath{P}|$ map (left) at 200 arcmin and the $B$-mode map (right) seen from the poles both overlaid with the gradient of $\bmath{P}$ maxima chains. The two top spherical projections shows the northern Galactic pole and the two bottom spherical projections shows the southern Galactic pole. \label{fig:Speric_high_lat}}
\end{figure*}

An exhaustive classification of all peculiar structures seen in $B$-type polarisation and $|\nabla \bmath{P}|$ as a function of angular scales for the entire southern hemisphere is beyond the scope of this paper. Figure \ref{fig:Speric_high_lat} shows an overall comparison between the $|\nabla \bmath{P}|$ map at 200 arcmin and the $B$-mode map seen from the poles, both overlaid with the gradient of $\bmath{P}$ maxima chains. A close inspection of these maps reveals multiple correlations between the maxima chains and the spin-2 decomposition of the polarisation map.

\citet{2016A&A...586A.141P} measured an asymmetry between the variances of the $E$- and $B$-mode maps of the infrared dust emission at 353 GHz. In order to highlight filaments present at high Galactic latitude, the authors filtered the map between multipoles $\ell=30$ and 300, equivalent to 36 to 360 arcmin, using a wavelet spline function. In the case of polarised dust emission, the polarisation is caused by the alignment of the dust grains with the magnetic field lines, which makes the orientation of the magnetic field more likely to be correlated with the overall matter distribution. For non-thermal synchrotron emission, instead of observing a general alignment of the magnetic field lines with filamentary structures, we observe sharp changes in the magnetic field direction, producing patterns in $B$-type polarisation and double-jump features in the $|\nabla \bmath{P}|$ map. In many cases, these features do not correlate with fluctuations in the polarisation intensity map, $|\bmath{P}|$, or with fluctuations in the total intensity map, Stokes I. In addition, the orientation of the synchrotron polarisation is subject to Faraday rotation, which is dependant on non-thermal electron density and the magnetic field strength. Consequently, the mix of intrinsic polarised emission and Faraday rotation along the line of sight makes the analysis of structure formation in the magnetic field difficult. However, the gradient of $\bmath{P}$ and the decomposition in $E$- and $B$-mode allows us to reveal peculiar regions where the magnetic field traced with synchrotron emission is organised in such a way that it also produces local asymmetries between $E$- and $B$-type fluctuations. Some asymmetries are localised at the edges of \HII regions or $H\alpha$ bubbles (Fig. \ref{fig:Eridanus}); others are located at high Galactic latitude without visible correlation with $H\alpha$ (Fig. \ref{fig:Speric_high_lat}). Such asymmetry, can often be described, but not always, as $B$-mode filamentary structures with the polarisation vector aligned $\sim45^{\circ}$ to the structure (Fig. \ref{fig:EB_lic}). The wavelet power spectra in Figures \ref{fig:pow_spec} and \ref{fig:excessB} show that those asymmetries can also occur at different angular scales, where the $E$- or $B$-type can dominate.

The best-fit power-laws for the two regions in Table \ref{tab:fit} respect the general tendency noticed by \citet{2010MNRAS.405.1670C} where fields at high Galactic latitude ($b=[-90^{\circ},-20^{\circ}]$) have steeper power-laws. The best-fit power-law for both $E$- and $B$-mode in the G353-34 field are close to their median value for this latitude range, $\beta^{(E+B)/2}=-2.6$ and the power-laws for the Orion-Eridanus field are close to those measured for low latitude fields ($b=[-20^{\circ},0^{\circ}]$). The authors explain this power-law variation by mid-high latitudes characterised by smooth emission with most of the power on large angular scales, and a disc field more evenly distributed with a slight predominance of power at smaller scales. The measured variance ratio $S_{B}/S_{E}$ for two different regions, one near the Galactic plane and one at lower Galactic latitude, varied between 1.04 and 1.43. \textit{Planck} results \citep{2015arXiv150201588P, 2016A&A...586A.133P, 2016A&A...586A.141P} have shown that an asymmetry of $S_{B}/S_{E} \approx 0.50$ exists between the amplitudes of power spectra fits of Galactic $B$- and $E$-modes in dust polarisation at 353 GHz. \citet{2015arXiv150201588P} also reported an $E$--$B$ asymmetry ($S_{B}/S_{E} \approx 0.35$) for the synchrotron emission at 30 GHz. These asymmetries are assumed to be caused by the magnetic field alignment with filamentary structures. For these studies, no $S_{B}/S_{E}$ ratio, above 1.0 were reported. Beside the fact that polarised synchrotron emission and polarised dust emission are produced by two different physical processes, synchrotron emission at 2.3 GHz is also affected by Faraday rotation. Most of the features associated with an excess are not visible in the polarisation intensity map, which means that these features are probably caused by spatial fluctuations in Faraday rotation only. Those perturbations of the magnetic field seem to be trigged, at least in part, by energy injected by star clusters associated with star forming regions or superbubbles. However, as it was demonstrated with the $B$-mode excess associated with Loop I in Fig. \ref{fig:excessB}, some excess in $E$- or $B$-mode can also be attributed to features seen in polarisation intensity.

The origin of the peculiar structures mentioned in this paper is diverse. It is generally accepted that the random and ordered magnetic field components can be roughly separated in terms of, respectively, small-scale turbulent fluctuations and large-scale coherent fluctuations aligned with the spiral arms and the structures in the Galactic halo. The latter component is nicely shown by the large-scale coherent structures associated with the giant magnetised outflows in Fig. \ref{fig:halo}. Turbulence certainly plays a major role in the structure formation of the random component of the magnetic field. Nevertheless, other physical processes occurring at smaller scales than the Galactic arms can produce extended coherent features such as \HII regions, interstellar bubbles or superbubbles. The close interplay between the ionised medium and magnetic field also increases the complexity of the magneto-ionic medium and challenges the analysis of structure formation in the ISM. Structures visible in $B$-modes in Fig. \ref{fig:EB_lic} are good examples of where a sharp deviation of the magnetic field lines is well correlated with the $H\alpha$ emission and where the origin of the structure is unclear. 

Consequently, simulations of the magneto-ionic medium, as a tool to characterise structures seen in observations or to build a model of the Galactic radio foreground emission, have to take into account the multiple origins and the great complexity of the medium in order to give a reliable description of the physical processes in action, and to create a representative model of the Galactic radio polarised emission.

\section{Conclusion}\label{sec:conclusion}

We have extended the previous analysis of \citet{2014A&A...566A...5I} of the spatial gradient of the synchrotron linear polarisation on the S-PASS survey at multiple spatial scales, following the method introduced by \citet{2015MNRAS.451..372R}. It is shown that the method can be easily adapted for spherical maps and that the multiscale technique can reveal different sets of filament-like networks previously filtered out with the original calculation of $|\nabla \bmath{P}|$.

We demonstrate that the rotationally invariant $\overline{\eth}^2 \eth^2\Psi_E$ and $\overline{\eth}^2 \eth^2\Psi_B$ fluctuations in the real space proposed by \citet{1997PhRvD..55.1830Z} have similar properties to applying the gradient directly on scalar fields $E$ and $B$ without the spin operators. We conclude that $|\nabla E|$ and $|\nabla B|$ are included in $|\nabla \bmath{P}|$ and that both analyses are complementary. This decomposition can be used to locate peculiar structures in synchrotron polarisation data and to help characterise the influence of nearby physical processes on structures in the magnetic field.

A more extended analysis of peculiar structures seen in the polarisation gradient and in the $E$- and $B$-mode maps of synchrotron emission could lead to a classification of different types of structure according to their spatial correlation with structures seen in polarisation intensity or $H\alpha$. By using similar methods to those introduced in the present paper, such classification could help us to identify the energy injection contribution of different kinds of physical processes, such as star formation regions or SNRs, in the Galactic magnetic field and measuring their influence in the nearby turbulent medium. New algorithms, such as the calculation of directional spin wavelet transforms on the sphere \citep{2015arXiv150906749M} and curvelets on the sphere \citep{2017ITSP...65....5C}, which allows the extraction of filamentary structures in a spin-2 signal, are promising avenues that will be explored in the future. Applying such analysis to MHD simulations could also allow us to experiment with the type of perturbations that influence the interstellar magnetic field and to compare their signature with observations.

\section*{ACKNOWLEDGEMENTS}

The authors gratefully acknowledge support from the European Research Council under grant ERC-2012-StG-307215 LODESTONE. This work has been carried out in the framework of the S-band Polarisation All Sky Survey (S-PASS) collaboration. The Parkes Radio Telescope is part of the Australia Telescope National Facility, which is funded by the Commonwealth of Australia for operation as a National Facility managed by CSIRO. The Dunlap Institute is funded through an endowment established by the David Dunlap family and the University of Toronto. B.M.G. acknowledges the support of the  Natural Sciences and Engineering Research Council of Canada (NSERC) through grant RGPIN-2015-05948, and of the Canada Research Chairs program. M.H. acknowledges the support of research programme 639.042.915, which is partly financed by the Netherlands Organisation for Scientific Research (NWO).

\bibliographystyle{mnras}
\bibliography{biblio}

\clearpage
\onecolumn

\appendix
\section{$|\nabla \bmath{P}|$ and $|\nabla E B|$ wavelet power spectra}\label{sec:appendix}

In the small-scale limit, since both decompositions are tracing all fluctuations of the polarisation vector $\bmath{P}$, the wavelet power spectrum of $|\nabla \bmath{P}|$ is equal to the wavelet power spectrum of $|\nabla EB|$. The definition of $|\nabla EB|$ in the wavelet space is

\begin{equation}
|\nabla \tilde{E}\tilde{B}(l,\bmath{x})| = \sqrt{ |\nabla \tilde{E}(l,\bmath{x})|^2 + |\nabla \tilde{B}(l,\bmath{x})|^2}
\label{A1}
\end{equation}

\noindent where,

\begin{equation}
\begin{split}
|\nabla \tilde{E}(l,\bmath{x})|^2 = |\tilde{E}_{1}(l,\bmath{x})|^2 + |\tilde{E}_{2}(l,\bmath{x})|^2,\\
|\nabla \tilde{B}(l,\bmath{x})|^2 = |\tilde{B}_{1}(l,\bmath{x})|^2 + |\tilde{B}_{2}(l,\bmath{x})|^2.
\end{split}
\label{A2}
\end{equation}

\noindent In the flat sky limit, the relation between the $E$-, $B$-mode and the $Q$, $U$ Stokes parameters can be defined in the Fourier domain through the rotation of the $k$-space \citep{1997ApJ...482....6S}:

\begin{equation}
\begin{split}
\hat{E}(\bmath{k}) &= \hat{Q}(\bmath{k})\cos(2\phi_{\bmath{k}}) + \hat{U}(\bmath{k})\sin(2\phi_{\bmath{k}}),\\
\hat{B}(\bmath{k}) &= -\hat{Q}(\bmath{k})\sin(2\phi_{\bmath{k}}) + \hat{U}(\bmath{k})\cos(2\phi_{\bmath{k}}),
\end{split}
\label{A3}
\end{equation}

\noindent where $\phi_{\bmath{k}}$ is the direction angle of the two-dimensional vector $\bmath{k}$. Because the wavelet power spectrum can also be defined directly in the the Fourier domain via the relation \citep{1992AnRFM..24..395F}

\begin{equation}
S(l) = \int \frac{|\tilde{f}(l,\bmath{x})|^2}{l^n} d\bmath{x} = \int |\hat{f}(\bmath{k})|^2|\hat{\psi}_{l}(\bmath{k})|^2 d\bmath{k},
\end{equation}

\noindent where $\hat{\psi}_{l}(\bmath{k})$ is the Fourier transform of the wavelet function $\psi(\bmath{x})$ with the scaling factor $l$, the wavelet power spectrum of the $E$-mode gradient in equation (\ref{A2}) can be defined as 

\begin{equation}
S_E(l) = \int |\hat{E}(\bmath{k})|^2|\hat{\psi}_{1l}(\bmath{k})|^2 + |\hat{E}(\bmath{k})|^2|\hat{\psi}_{2l}(\bmath{k})|^2 d\bmath{k}
\label{A4}
\end{equation}

\noindent where $\hat{\psi}_{1,2l}(\bmath{k})$ are the Fourier transform of the DoG wavelets. From equations (\ref{A2}), (\ref{A3}) and (\ref{A4}):

\begin{equation}
\begin{split}
S_E(l) &= \int |\hat{Q}(\bmath{k})\cos(2\phi_{\bmath{k}}) + \hat{U}(\bmath{k})\sin(2\phi_{\bmath{k}})|^2|\hat{\psi}_{1l}(\bmath{k})|^2
		+|\hat{Q}(\bmath{k})\cos(2\phi_{\bmath{k}}) + \hat{U}(\bmath{k})\sin(2\phi_{\bmath{k}})|^2|\hat{\psi}_{2l}(\bmath{k})|^2 d\bmath{k},\\
	      &= \int |\hat{Q}_{1l}(\bmath{k})|^2\cos^2(2\phi_{\bmath{k}}) +  |\hat{U}_{1l}(\bmath{k})|^2\sin^2(2\phi_{\bmath{k}})
	      + 2\times|\hat{Q}(\bmath{k})\hat{U}(\bmath{k})\cos(2\phi_{\bmath{k}})\sin(2\phi_{\bmath{k}})||\hat{\psi}_{1l}(\bmath{k})|^2\\
	      &\quad+ |\hat{Q}_{2l}(\bmath{k})|^2\cos^2(2\phi_{\bmath{k}}) +  |\hat{U}_{2l}(\bmath{k})|^2\sin^2(2\phi_{\bmath{k}})
	      + 2\times|\hat{Q}(\bmath{k})\hat{U}(\bmath{k})\cos(2\phi_{\bmath{k}})\sin(2\phi_{\bmath{k}})||\hat{\psi}_{2l}(\bmath{k})|^2d\bmath{k},
\end{split}
\label{A5}
\end{equation}

\noindent where $|\hat{Q}_{1,2l}(\bmath{k})|^2=|\hat{Q}(\bmath{k})|^2|\hat{\psi}_{1,2l}(\bmath{k})|^2$. Similarly for $B$-mode gradient,

\begin{equation}
\begin{split}
S_B(l) &= \int |-\hat{Q}(\bmath{k})\sin(2\phi_{\bmath{k}}) + \hat{U}(\bmath{k})\cos(2\phi_{\bmath{k}})|^2|\hat{\psi}_{1l}(\bmath{k})|^2
		+|-\hat{Q}(\bmath{k})\sin(2\phi_{\bmath{k}}) + \hat{U}(\bmath{k})\cos(2\phi_{\bmath{k}})|^2|\hat{\psi}_{2l}(\bmath{k})|^2 d\bmath{k},\\
	      &= \int |\hat{Q}_{1l}(\bmath{k})|^2\sin^2(2\phi_{\bmath{k}}) +  |\hat{U}_{1l}(\bmath{k})|^2\cos^2(2\phi_{\bmath{k}})
	      -2\times|\hat{Q}(\bmath{k})\hat{U}(\bmath{k})\cos(2\phi_{\bmath{k}})\sin(2\phi_{\bmath{k}})||\hat{\psi}_{1l}(\bmath{k})|^2\\
	      &\quad+ |\hat{Q}_{2l}(\bmath{k})|^2\sin^2(2\phi_{\bmath{k}}) +  |\hat{U}_{2l}(\bmath{k})|^2\cos^2(2\phi_{\bmath{k}})
	      -2\times|\hat{Q}(\bmath{k})\hat{U}(\bmath{k})\cos(2\phi_{\bmath{k}})\sin(2\phi_{\bmath{k}})||\hat{\psi}_{2l}(\bmath{k})|^2d\bmath{k}.
\end{split}
\label{A6}
\end{equation}

\noindent Consequently,

\begin{equation}
\begin{split}
S_{\tilde{E}\tilde{B}}(l) &= S_E(l) + S_B(l)\\
		&=\int |\hat{Q}_{1l}(\bmath{k})|^2 + |\hat{Q}_{2l}(\bmath{k})|^2 + |\hat{U}_{1l}(\bmath{k})|^2 + |\hat{U}_{2l}(\bmath{k})|^2 d\bmath{k},\\
		&=S_{\tilde{\bmath{P}}}(l),
\end{split}
\label{A7}
\end{equation}

\noindent where $S_{\tilde{E}\tilde{B}}(l)$ and $S_{\tilde{\bmath{P}}}(l)$ are respectively the wavelet power spectra of $|\nabla \tilde{E}\tilde{B}(l,\bmath{x})|$ and $|\nabla \tilde{\bmath{P}}(l,\bmath{x})|$.

\bsp
\label{lastpage}
\end{document}